\documentclass[reqno,12pt]{amsart}
\usepackage{hyperref}
\usepackage{amssymb,amsmath,amsthm}
\usepackage{a4wide}

\usepackage{color}
\usepackage{esint}
\usepackage{mathrsfs}
\usepackage{calrsfs}
\usepackage[toc,page]{appendix}
\makeatletter
\renewcommand \theequation {%
\ifnum \c@section>\z@ \@arabic\c@section.%

\fi\@arabic\c@equation} \@addtoreset{equation}{section}

\makeatother

\theoremstyle{definition}

\theoremstyle{remark}

\def\XXint#1#2#3{{\setbox0=\hbox{$#1{#2#3}{\int}$ }
\vcenter{\hbox{$#2#3$ }}\kern-.6\wd0}}

\providecommand{\abs}[1]{\left\vert#1\right\vert}

\providecommand{\nnm}[1]{\left\vert\kern-0.25ex\left\vert\kern-0.25ex\left\vert#1\right\vert\kern-0.25ex\right\vert\kern-0.25ex\right\vert}

\def\ud{\mathrm{d}}

\def\p{\partial}

\def\r{\mathbb{R}}

\def\ds{\displaystyle}

\def\ab{\overline{\mathscr{A}}}
\def\a{\mathscr{A}}
\def\x{\xi}
\def\tt{\tau}

\def\e{\varepsilon}

\def\vx{x}
\def\vv{v}
\def\nx{\nabla_{x}}

\def\d{\delta}

\def\G{R\!\!\!\!\!\kern-1.ptI\,\,}
\def\TT{T\!\!\!\!\kern-1.1pt I\,\,}
\def\RR{R\!\!\!\!\!\kern-1.ptI\,\,}

\def\lc{\mathcal{L}}

\def\li{\lc^{-1}}

\def\a{\mathscr{A}}
\def\ab{\bar{\mathscr{A}}}
\def\m{\mu}

\def\be{\mathbf{e}}

\def\rq{\rho}
\def\uq{u}
\def\tq{T}

\def\k{\kappa}

\def\b{\mathscr{B}}
\def\bbb{\bar\b}

\def\la{\lambda}

\def\pt{\partial}
\def\l{\lambda}

\def\e{\varepsilon}
\def\a{\alpha}
\def\be{\beta}
\def\d{\delta}

\def\newpage{\vfill\eject}

\def\to{\rightarrow}

\def\frac{\over}
\def\\{\cr}
\def\ref{}
\def\nonumber{}

\vfill

\def\ga{\gamma}
\def\a{\alpha}

\def\d{\delta}

\def\z{\zeta}

\def\sneq{=\hskip-.18cm/\hskip.07cm}

\numberwithin{equation}{section}

\title[On the derivation of new non-classical hydrodynamic equations\dots]{On the derivation of new non-classical hydrodynamic equations for Hamiltonian particle systems. }

\author[R. Esposito]{Raffaele Esposito}
\address[R. Esposito]{
   \newline\indent M\&MOCS - Universita' dell'Aquila}
\email{raff.esposito@gmail.com}

\author[R. Marra]{Rossana Marra}
\address[R. Marra]{
   \newline\indent Dipartimento di Fisica and Unit\`a INFN, Universit\`a di Roma Tor Vergata}
\email{marra@roma2.infn.it}
\thanks{R. Marra is supported by INFN}

\begin{document}
\begin{abstract}
We consider a Hamiltonian   system of particles, interacting through of a smooth  pair potential. We look at the system on a space scale of order $\e^{-1}$, times of order $\e^{-2}$, and mean velocities of order $\e$, with $\e$ a scale parameter, under initial conditions where the system is in a local Gibbs state with parameters corresponding to density and temperature with gradients of order $1$. Assuming that the phase space density of the particles is given by a suitable series in $\e$  the behavior of the system under this rescaling is described, to the lowest order in $\e$, by new non-classical  hydrodynamic equations that cannot be derived from the compressible Navier-Stokes equations in the small Mac number limit. The analogous equations in kinetic theory are called ghost effect equations.

\end{abstract}

\maketitle

\vskip .5cm
\section{ Introduction}

The problem of deriving the  hydrodynamical
equations from the Hamiltonian equations of motion of atoms, in the limit when a scale parameter $\e$ is small, is one of the main open problems of non-equilibrium Statistical Mechanics.
The compressible Navier-Stokes system (CNSE) is  a phenomenological  description of the  dissipative hydrodynamics and the incompressible Navies-Stokes-Fourier system (INSF) can be derived  from the compressible one in the low Mach number limit. Unfortunately the CNSE has no scaling  space-time invariance and hence cannot be obtained from a microscopic description while such an obstruction is not present for the INSF system and, assuming the temperature constant at the $\e^0$ order, a formal derivation from particles was given in  \cite{EM}, while a rigorous proof at the level of the Boltzmann equation was obtained, among the others, in \cite{DEL}, \cite{EGKM}. 

The situation is very different when the assumption of constant temperature at $\e^0$ order is removed. In this case, a set of new physically relevant hydrodynamical equations has been derived formally starting from the Boltzmann equation \cite{So}\cite{Ko}\cite{KGF}\cite{DEL}\cite {Bo}. They are characterized by  a correction to the Navier-Stokes stress tensor that depends on derivatives of the temperature. This effect was already known to Maxwell \cite{Max}.  
The relevance of this system  is that  it is non-classical in the sense that it cannot be derived from the CNSE, and hence in some sense it indicates a failure of the CNSE in describing the real
world.  Since these equations are derived from the Boltzmann equation, the state equation is the one of a perfect gas.
Sone has given to these equations the name of ghost effect system. The name is suggested by the fact that a vanishingly small velocity field produces finite size modifications of the usual heat equation. There are situations (particular geometries, stationary cases etc.) in which the classical heat-conduction equation fails to correctly describe the temperature field of the gas.  These modifications are confirmed
by many  numerical experiments. There has been a big theoretical, numerical and
experimental work on this and for the details we refer to \cite{So} and references therein. 

It is a natural question to ask if such kind of equations can be derived also 
from a Hamiltonian particle system. In this paper we answer affirmatively to that, but only at a formal level.

 A  system of many interacting particles, moving according to 
the Newton equations of motion, can be described on a space 
scale much larger than the typical microscopic scale
(say the range of the interaction) in terms of 
density, velocity and temperature fields, satisfying
hydrodynamic equations, like Euler or Navier-Stokes equations. 
The scale separation and the local conservation laws are
responsible of this reduced description.

In fact, on the macroscopic scale the quantities which are locally 
conserved (slow modes) play a major role in the motion of the fluid.
The derivation of the Euler equations is based on the assumption of 
local equilibrium. On times of order 
$\e^{-1}$, the system is expected to be described approximately
by a local Gibbs measure, with parameters varying
on regions of order $\e^{-1}$, $\e$ being a scale parameter. 
The local equilibrium assumption implies that the parameters 
of the local Gibbs measures satisfy the Euler equations
\cite{M}, \cite{De M}, \cite{DP}.
The microscopic structure (the potential) appears only in 
the state equation which links  pressure and internal energy 
to the other 
macroscopic parameters. The microscopic locally conserved
quantities converge, as $\e \to 0$, by a law of large numbers, 
to macroscopic fields.
To make this correct, the many particles Hamiltonian 
system must have good dynamical mixing properties to 
approach and stay in a  state close to the local equilibrium. 
At the moment it is not understood how to provide such properties. 
Therefore the only rigorous results are obtained
by adding some noise to the Hamiltonian evolution \cite{OVY}
(see \cite{S}  for a review on the rigorous results for
stochastic systems).
 
The derivation 
of the Navier-Stokes equations presents many more difficulties. These equations, which 
describe the behavior of a fluid in the presence of
dissipative effects, do not have an immediate interpretation 
in terms of scale separation. This is not surprising because
the NS equations do not have a natural space-time scale
invariance like the Euler equations. In fact to see the
effect of the viscosity and the thermal conduction one has to
look at times such that neighboring regions in local equilibrium
exchange a sensible amount of momentum and energy. Simple
considerations show that the right scale of time is 
$\e^{-2}$. On the other hand we cannot hope to find the compressible
Navier-Stokes-Fourier  behavior from the particle system under the parabolic rescaling $x \to \e^{-1}x$ and 
$t \to \e^{-2}t$ since the NS equations are not invariant 
under this scaling, due to the presence of the transport terms.
A way out is to consider the incompressible limit simultaneously,
because the incompressible Navier-Stokes-Fourier equations (INSF)
have the required scaling invariance. 
 Along this path, in \cite{EM} we gave a
formal derivation of the INSF from a Hamiltonian particle system
under the parabolic rescaling, in the low Mach number regime.

In the paper \cite{EM} the main ingredient is the assumption that the non-equilibrium 
density solution of the rescaled Liouville equation can be expressed as a truncated series in the parameter
$\e$.  We followed a procedure inspired to the Hilbert-Chapmann-Enskog
expansion \cite{Ce}, used to construct the  solution of the
rescaled Boltzmann equation. From the physical point of view we
think of the system as being in local equilibrium  with
parameters which are themselves given by a series in $\e$. 
However there is a non-hydrodynamic
correction to the local equilibrium which depends  from the non
conserved quantities in the system (fast modes) and we assume
that this correction does not affect the  first order in the
expansion, that is at the first order the system is still
described by a local Gibbs measure with parameters  which differ
from constants by terms of order $\e$. The parameters conjugate to density and temperature are constant plus terms of order $\e$ and the one conjugate to the velocity field is of order $\e$. This last is strictly
related to the incompressibility assumption and would be false in
the case of finite Mach number. 
This assumption is the translation of the Hilbert expansion for the Boltzmann equation  to the particle system
case. 
On the other hand, the non hydrodynamical corrections in the 
second order are important on the scale $\e^{-2}$ and give rise to 
the N.S. terms.  
It is worth to mention here that very strong and rather 
uncontrollable assumptions are necessary
even to give sense to the formal calculations below: 
\begin{itemize}
\item the space of the invariant observables
for the microscopic dynamics reduces to the locally
conserved quantities, mass, momentum and energy and functions of them;
\item some equilibrium time correlation functions decay 
sufficiently fast.
\end{itemize}
\noindent Such assumptions are far from being sufficient for 
a mathematical proof.

Under the same scaling, in the context of the Boltzmann equations, for initial conditions such that the density and temperature have gradients of order $1$ the formal limiting equations are different  and some new terms depending on gradients of temperature appear in the momentum equations and  the velocity field is not anymore divergence-less, see e.g.  \cite{So}\cite{Ko}\cite{DEL}. 
The rigorous derivation for the stationary Boltzmann equation has been obtained recently in \cite{EGMW}.

Here, we try to formally derive the analogous equations as limiting equations from a system of interacting particles. 
Again, the main ingredient is the assumption that the non-equilibrium 
density can be expressed as a truncated series in the parameter
$\e$. The difference, to be consistent with the initial conditions, is that the first term of the expansion is the Gibbs measure with parameters conjugate to density and temperature depending on time and position.

This a particular {\it local} Gibbs measure and we denote it by $G_0$. We stress that it is not anymore an equilibrium.

If the initial condition is the local Gibbs measure  $G_0$, plus correction of order $\e$, the empirical fields evolve on the macroscopic space-time scales close to the solution of the non-classical new equations, while
if the initial condition is the global Gibbs measure plus correction of order $\e$, the empirical fields evolves on the macroscopic space-time scales close to the solution of the INSF,  since in this case the perturbation of order $\e$ is too small to generate the new terms in the limiting equations.  

In Section 2 we introduce the empirical fields and the microscopic evolution equations for them in terms of the currents. The currents  cannot be expressed back in terms of the hydrodynamical fields, and  one has to solve the closure problem in a suitable way to get the hydrodynamical equations. We consider the averages versus $F^\e$, solution of the Liouville equation, and the lowest order in $\e$ gives the following hydrodynamic equations, derived in Section 3, for the velocity field $u$, the 
density $\rho$ and the internal energy $e$ 

\begin{align}\label{fluid system-1}
\left\{
\begin{array}{rcl}
    \nabla P&=&0,
    \\
    \rule{0ex}{1.2em}    {\partial_ t}\rho+\nx\cdot(\rq\uq)&=&0,
    \\
    \rule{0ex}{1.2em}
   \rho\partial_t\uq+\uq\cdot\nx\uq+\nx \mathfrak{p}&=&\nx\cdot\left(\tau^{(1)}-\tau^{(2)}\right),
   \\
   \rule{0ex}{1.8em}
\rho[ \partial_t e+ u\cdot\nabla e]+P\big(\nx\cdot\uq\big) &=&\nx\cdot\left(\k\dfrac{\nx\tq}{2\tq^2}\right),
    \end{array}
    \right.
\end{align}
where, for  $\a,\beta=1,\dots d$, with $d$ the space dimension,
$$\tau^{(1)}_{\a\beta}:=\eta\left(\p_\a u^{\beta}+\p_\beta u^{\a}-{2\over d}\d_{\a\beta}\p_\a u^{\a}\right)+\zeta \d_{\a\beta}\p_\a u^{\a} ,$$
$$\tau^{(2)}_{\a\beta}:=[K_1({\partial_\a }T\partial_\beta T-{1\over d}\sum_\alpha({\partial_\a }T)^2) +{\omega_1}  \sum_\alpha({\partial_\a }T)^2+ K_2 (\partial^2_{\a\beta}T-{1\over d}\sum_\alpha\partial^2_{\a\alpha}T)+{\omega_2}\sum_\alpha\partial^2_{\a\alpha}T)]$$  
Here, $P$ and $e$ are the thermodynamical functional pressure and internal energy which are functions of  $\rho, T$ as determined by the Gibbs    measure.

These equations are a new set of hydrodynamic equations, which cannot be derived from the compressible Navier-Stokes equations. The divergence of the velocity field  is not zero as in INSF, even if we sent the Mach number to zero. There are new "thermal stress" terms in the equation for the velocity field. We observe that, even if in the thermal stress tensor $\tau^{(2)}$ there are third order derivatives in $T$, these terms (the last two) appear in the equation as a gradient and hence can be absorbed in the unknonw pressure $\mathfrak{p}$ \cite{Le}\cite{So}.

We also show that the entropy associated to the solutions of these equations, correctly, is increasing in time.

Our derivation gives the viscosity coefficient $\eta$, bulk viscosity
$\zeta$ and the conductivity $\k$ 
in terms of the local  equilibrium time correlation functions.

The fluctuation-dissipation
theory relates the transport coefficients to time-integrated
correlation functions (see \cite {G}, \cite {K}). 
The expression
we find agrees with the Green-Kubo formula for the shear and bulk viscosity as well for the conductivity.  
 Moreover, in these new equations, in particular in the equation for the momentum,  there are more transport coefficients and we find an expression for them in terms of the potential and of a double time integral in the Appendix. These expression reminds the ones in the second order dissipative hydrodynamics introduced in \cite{Z} (\cite{HSR}). We remark that 
 the linear response theory cannot give these new transport coefficients.
 
 In the last section we compare this equations with the ones derived from the Boltzmann equation.

\vskip .5cm

\section{Conservation laws} 

\vskip .5cm
\noindent 
We consider a system of many, $N$, identical  particles of unit mass  in a
torus $\Lambda_\e$ of size $\e^{-1}$ in $\Bbb R^d$,   
interacting via a pair central potential $V$ of finite range. 
The Newton equations are
\begin{equation*}
\begin{split} {d\x_i\over d\tt}(\tt) &= v_i(\tt), \notag\\
{dv_i\over dt}(\tt)&= - \sum_{i \ne j} \nabla 
V(\x_i  
-\x_j |) ,\end{split}
\end{equation*}
where $\xi_i, v_i, \tt$, denote the microscopic coordinates, velocities and time, and  $i=1,\dots,N$, 
After rescaling space as $\e^{-1}$ and time  as $\e^{-2}$, they become,
\begin{equation}
\begin{split} {dx_i\over dt}(t) &= \e^{-1}v_i(t), \\
 \label{(2.1)}\\
{dv_i\over dt}(t)&= - \e^{-2}\sum_{i \ne j} \nabla 
V(\e^{-1}|x_i  
-x_j |) , \end{split}
\end{equation}
 with $x_i=\e\xi_i$  the macroscopic coordinates and $t=\e^2 \tau$ the macroscopic time.
The number of particles $N$ is assumed to be of order  $\e^{-d}$
to keep the density finite. 

The  rescaled Newton equations, with initial data  randomly distributed,  are equivalent to the Liouville equation for the evolution of a distribution $\mu^\e_0(x_1,\cdots,x_N,v_1,\cdots,v_N):=\mu_0(\e^{-1}x_1,\cdots,\e^{-1}x_Nv_1,\cdots,v_N)$ at time $0$ on the $N$ particle  phase space 
\begin{equation}{\partial\over\partial t}\mu^\e_t={\e^{-2}}™{\mathcal L}^*\mu^\e_t, \label{liou}\end{equation}
where ${\mathcal L}^*$ is the Liouville  operator 
\begin{align}
{\mathcal L}^* \phi(x_1,\cdots,x_N,v_1,\cdots,v_N) = -\sum_{i=1}^N \sum_{k=1}^d
\big\{\e v_i ^k {\partial \phi\over \partial x_i ^k} -
\sum_{i \sneq j}  {\partial_ k} V\Big(\e^{-1}|x_i  
-x_j |\Big) {\partial \phi \over \partial v_i ^k }
\big\},\label{(3.17)}
\end{align} 
with ${\mathcal L}^ *=-{\mathcal L}$ the adjoint of ${\mathcal L}$ with respect
to the scalar product induced by the a priori measure $d \mathbb{Z}={1\over n!}d^d x_1 d^d v_1$  $\dots d^d
 x_N d^d v_N$. To be more explicit,
 we  denote the average versus the a priori measure 
 $d \mathbb{Z}$ by $\Big\langle \phi\Big\rangle$ and  define the scalar
  product
$\langle \phi,\psi\rangle=\Big\langle \phi\psi\Big\rangle$
so that
$$\langle \phi,{\mathcal L}^*\psi\rangle=-\langle{ \mathcal L}\phi,\psi\rangle.$$
 The total number of particles, the
$d$ components of the total momentum and the  total energy are the
conserved quantities. We define the corresponding empirical
fields:
\vskip.1cm

\noindent {\it empirical density} 
\begin{equation}z^0 (x) = \e^d \sum_{i=1}^N \delta (x_i -x), \label{(2.2)}\end{equation}
{\it empirical velocity field density}
\begin{equation}z^{\a}(x) =\e^d \sum_{i=1}^N v^{\a} _i \delta (x_i -x), 
\phantom{...} \a =1,\dots,d , \label{(2.3)}\end{equation} 
{\it  empirical energy density}

\begin{equation}z^{d+1}(x) =\e^d \sum_{i=1}^N{1\over 2}\big[ v^2 _i +\sum_{j \ne i=1}^N
 V(\e^{-1}|x_i -x_j|)\big] \delta (x_i -x)] .\label{(2.4)}\end{equation}
Their meaning is as follows: The average of the integral of  
$z^{\a}$ over a small region  is equal to the
average number of particles, momentum, energy associated  to the
region. We  will write use the short notation 
\begin{equation}z^{\mu}(x)=\e^d \sum_{i=1}^N \delta (x_i -x) z_i ^\mu,  \label{(2.5)}
\end{equation}
with 
$$z_i ^0 =1;\phantom{..} z_i ^\a =v_i ^\a, \a=1,\dots,d;
\phantom{..} z_i ^{d+1} = {1\over 2}[ v^2 _i +\sum_{i \ne j=1}^N
V(\e^{-1}|x_i -x_j|)].$$
Note that the correct way of writing these quantities would be to add an $\e$ index: 
$z_\e^\mu(x)=z^\mu(\x)$. We will omit this index in the following, being clear from the context whether we are referring to microscopic or macroscopic variables.

The empirical fields satisfy the following local conservation 
laws, which are obtained by integrating \eqref{(2.5)}  against a smooth $\Bbb R^d$ test function $f(x)$,  $t$-differentiating and using the Newton equations \eqref{(2.1)}:
\begin{equation}{d\over dt}\e^d \sum_{i=1}^N f(x_i)=\e^{-1} \e^d \sum_{i=1}^N\sum_{\a=1}^d
{\partial_\alpha f}(x_i) v_i^\a ,  \label{(2.7)}\end{equation}
\begin{equation} {d\over dt} \e^d \sum_{i=1}^Nf(x_i)v_i^\be =
 \e^{-1} \e^{d} \sum_{i=1}^N \sum_{\a=1}^d\Big\{
{\partial_\alpha f}(x_i) v_i^\a v_i^\be -\e^{-1}  
\sum_{j \ne i=1}^N
 {\partial_\beta} V (\e^{-1}|x_i -x_j|) f(x_i) \Big\} , \label{(2.8)}\end{equation}
\begin{equation} {d\over dt} \e^d \sum_{i=1}^N f(x_i) z_i^{d+1} = \e^{-1} \e^d
\sum_{i=1}^N \sum_{\a=1}^d \Big\{ {\partial_\a f}(x_i) v_i^\a
z_i^{d+1}   -{1 \over 2}\e^{-1}\sum_{i \ne j=1}^N
 {\partial_\a}  V (\e^{-1}|x_i -x_j|) v_i^\a f(x_i)  \Big\}. \label{(2.9)}\end{equation}
Here $\partial_\beta V(\xi)= \partial V(\xi)/\partial \xi_\beta$. 
Because of the symmetry properties of the potential we can write, 
as usual, the second term in the r.h.s. of \eqref{(2.8)} as
\begin{equation}-{1 \over 2}\e^{d-2}\sum_{i \ne j=1}^N
 {\partial_\beta} V (\e^{-1}|x_i -x_j|)[f(x_i) -f(x_j)] . \label{(2.10)}\end{equation}
Since $f$ is slowly varying on the microscopic scale we
can write
\begin{equation}f(\e \xi_i) -f(\e \xi_j)= \sum_{\ga=1}^{d}{\partial_\ga f 
}(x_i) \e[\xi_i ^\ga  - \xi_j ^\ga]
+\e^2 D_0 + \e^3 D + O(\e^4) ,\label{(2.11)}
\end{equation}
where
\begin{equation} D_0={1\over 2}\sum_{\ga,\nu=1}^{d}{\partial_{\a \nu}^2 }f (x_i) [\xi_i ^\ga  -\xi_j ^\ga ]
[\xi_i ^\nu  -\xi_j ^\nu ],\end{equation}
\begin{equation}D={1\over 6}\sum_{\ga,\nu,\a=1}^{d}{\partial_{\ga\a\nu}^3 f
 }(x_i) [\xi_i ^\ga 
-\xi_j^\ga ] [\xi_i ^\nu  -\xi_j ^\nu ] [\xi_i ^\a  -\xi_j^\a ].\end{equation}

The last term of \eqref{(2.8)} becomes:
\begin{equation}\begin{split}&{1 \over 2}\e^{-1} \e^d \sum_{i,j=1}^N\sum_{\ga=1}^{d}{\partial_\ga f
}(x_i) \Psi^{\be \ga}(\e^{-1}|x_i -x_j|)
\\+&
{1 \over 4} \e^d \sum_{i,j=1}^N\sum_{\ga,\nu=1}^{d}{\partial_{\ga\nu}^2 f
 }(x_i) \Phi_0^{\beta\nu\gamma}(\e^{-1}|x_i -x_j| )
\\+&
{1 \over 12}\e \e^d \sum_{i,j=1}^N\sum_{\ga,\nu,\a=1}^{d}{\partial_{\ga\nu\a}^3 f
}(x_i) \Phi^{\beta\a\gamma\nu}(\e^{-1}|x_i -x_j|)+O(\e^{2}).\label{(2.12)} \end{split}\end{equation} 
with 
\begin{equation}\Psi^{\be \ga}(\xi)= - \partial_\be V(\xi)\xi^{\ga},\quad \Phi_0^{\beta\nu\gamma}=- \partial_\be V(\xi)\xi^{\ga}\xi^{\nu},
\quad \Phi^{\beta\a\gamma\nu}(\xi)= - \partial_\be V(\xi)\xi^{\ga}\xi^{\a}\xi^{\nu}.\label{(2.12.1)}\end{equation}
We also set
\begin{equation}\begin{split}&\bar\Phi_0^{\beta\gamma\nu}(x)={1\over 2}\e^d \sum_{ij=1}^N \delta(x_i -x)\Phi_0^{\beta\gamma\nu}(\e^{-1}|x_i -x_j| ),\\&\bar\Phi^{\a\beta\gamma\nu}(x)={1\over 2}\e^d \sum_{ij=1}^N \delta(x_i -x)\Phi^{\a\beta\gamma\nu}(\e^{-1}|x_i -x_j|),\label{Phi}
\end{split}\end{equation}
An analogous computation, which we omit for sake of shortness, can be done for the energy equation.  

The general form of the rescaled local conservation laws is (up to higher order in $\e$)
\begin{equation}{\partial\over \partial t}\int dx f(x) z^\be (x) = \e ^{-1}
\int dx \sum_{k=1}^d {\partial_k f}(x) w^{\be
k}(x) ,\label{(2.13)}\end{equation}
where $w^{\be k}$, $\be=0,\dots,d+1;\ k=1,\dots,d$ are
the currents corresponding to the fields $z^\be$ and are explicitly
given by 
\begin{equation}w^{0k} (x) = \e^d \sum_{i=1}^N \delta(x_i -x) v_i ^k,
 \label{(2.14)}\end{equation}
\begin{align}
\begin{split}w^{\be k }(x) &= \e^d \sum_{i=1}^N \delta(x_i -x)  \big\{
v_i^\be v_i^k +  {1 \over 2}\sum_{j=1}^N
\Psi^{\be k} (\e^{-1}|x_i -x_j|) \big\}\\
&+{1 \over 4}\e\sum_{\ga=1}^{d}{\partial_\ga 
} \Phi_0^{k\beta\gamma }(x)
 +{1 \over 12}\e^2\sum_{\ga,\nu=1}^{d}{\partial_{\ga\nu}^2 
 } \Phi^{k\beta\gamma \nu}(x)+O(\e^{3}),
\quad \beta=1, \dots , d, \\
w^{d+1\, k}(x) &= \e^d \sum_{i=1}^N\big\{
 v_i^k z_i^{d+1} + {1 \over 2}\sum_{j=1}^N\sum_{\ga=1}^{d}
\Psi^{\ga k} (\e^{-1}|x_i -x_j|) {1 \over 2} [v^\ga _i +  v^\ga
_j]\big\} +O(\e^{}), \label{(2.145)}
\end{split}\end{align}
We  introduce the notation $w^{ \be k}_i$ such that $$w^{ \be k}(x)=\e^d \sum_{i=1}^N \delta(x_i -x)  w^{ \be k}_i.$$
Hence $w^{0k}_i=v_i^k$, and so on.

We also denote the r.h.s. of the first line of \eqref{(2.145)} by 
\begin{equation}w_*^{\be k }(x) := \e^d \sum_{i=1} ^N\delta(x_i -x)  \big\{
v_i^\be v_i^k +  {1 \over 2}\sum_{j=1}^N
\Psi^{\be k} (\e^{-1}|x_i -x_j|) \big\}.\label{w*}\end{equation}
We stress that we keep  terms of order $\e$ and $\e^2$ in \eqref{(2.145)} while we do not do the same in \eqref{w*} and that such new terms were not taken into account in the derivation of the incompressible Navier-Stokes  equation \cite{EM} because do not give contribution at the lowest order. In this new setting instead we will see that the $D$ term gives a contribution at the lowest order for the momentum equation.

A key point in the argument is the remark that the empirical fields
$z^\alpha(\xi)$ are {\it approximate integrals  of the motion} in the
following sense:

 \noindent For any smooth function $f(x)$ on the torus 
$\Bbb T_1^d$, from previous calculation   it follows 
 that
\begin{equation}\mathcal{L}\big[ \e^d\sum_{i=1}^N f(\e\xi _i) z_i^\a\big]
=O(\e). \label{(2.18)}\end{equation}

\noindent On the other hand, 
the evolution of a generic observable $\Phi(x_1,\cdots,x_N,v_1,\cdots,v_N)$ according to the
rescaled Newton equations is  given by
\begin{equation}\pt_t \Phi=\e^{-2}{\mathcal L}\Phi.\label{(ip)}
\end{equation}
\noindent
Therefore, while a generic observable has a time derivative of order
$\e^{-2}$, there are special observables as the empirical fields associated to the mass, momentum and
energy, that have a time derivative only of order $\e^{-1}$. So they have
a comparatively slow evolution and this justifies the denomination of
approximate integrals of motion we used above. 

We stress that we {\it need to  assume} that
the only  observables with this property are the empirical fields (and of course  any function of them as well).

\bigskip
Now we turn to the definition of a special class of states, the local Gibbs states. A (global) Gibbs state in a  macroscopic volume $\Lambda$ (hence a volume $\e^{-1}\Lambda$ 
in microscopic variables) is defined as a probability
distribution on the
$\G_\Lambda$-space $\bigcup_{N\ge 0}(\Lambda\times \Bbb R^d)^N$, 
whose density with respect to the a priori measure $d\mathbb{Z}$
is given by (in the grandcanonical setting)
\begin{align}&G_\Lambda(x_1,v_1,\dots,x_N,v_N)=
Z_\Lambda^{-1}\prod_{j=1}^N {\tilde z} \exp\Big[-{(v_j- u)^2\over 2T}
+{1\over T}\sum_{k\sneq
j}V\Big({| x_j- x_k|\over
\e}\Big)\Big].\label{(5.3.8)}
\end{align}

Here $\tilde z$ is the activity and $Z_\Lambda$ is a normalization factor. We can write the previous expression in terms of  $\underline{\lambda}=\{\l^\a\}$, $\a=0,\dots,d+1$, $d+2$  real 
numbers called {\it chemical potentials} associated to the mass, momentum and energy
empirical fields and parametrizing the family of the Gibbs states
\begin{equation}
G(\underline{\lambda})=  Z_\Lambda^{-1}\exp \{ \sum_{i=1}^N\sum_{\mu=0}^{d+1} \lambda ^\mu z_i ^\mu
\}.
 \end{equation}
 Their explicit expression is 
$$
\lambda^0=\log \tilde z-{1\over 2}\beta |u|^2,\quad \lambda^\nu={1\over T} u^\nu,\nu=1,\cdots,d\quad\lambda^{d+1}=-{1\over T}.
$$
A more
conventional parametrization is given by $\rho>0$, the {\it mass density}, $u\in
\RR^d$, the {\it velocity field} and 
$T>0$, the {\it temperature}. Note that for a global Gibbs state, due to the Galilean invariance, the velocity field is usually assumed to vanish. In the definition of local Gibbs states, on the contrary we keep a non vanishing velocity field.

The notion of {\it local} Gibbs state is related to the one of local equilibrium and is obtained formally by replacing the constant chemical potentials
$\{\l^\a\}$ by some smooth functions
$\{\l^\a(x,t)\}$ which are functions of $x_i=\e\xi_i$: 

hence the local equilibrium state is defined, for
a fixed $\e$ as
\begin{equation}
\tilde G(\underline{\lambda})=  \tilde Z_\Lambda^{-1}\exp \{ \sum_{i=1}^N\sum_{\mu=0}^{d+1} \lambda ^\mu(x_i,t) z_i ^\mu
\},
 \end{equation}
 or in the equivalent form
 \begin{equation}
 \tilde G(\underline{\lambda})= {  \tilde Z_\Lambda^{-1}}\exp
\left[\int_{\e^{-1}\Bbb T^d} d\xi
\sum_{\a=0}^{d+1}\l^\a(x,t)z^\a(\xi)\right]={  \tilde Z_\Lambda^{-1}}\exp
\e^{-d}\left[\int_{\Bbb T^d} dx
\sum_{\a=0}^{d+1}\l^\a(x,t)z^\a(\e^{-1}x)\right],\label{(lgs)}
\end{equation}
with ${  \tilde Z_\Lambda}$ a suitable normalization factor and ${\Bbb T^d} $ the macroscopic torus.

 Note  that the local Gibbs states corresponding to any choice of smooth
chemical potentials, being given by the exponential of linear
combinations of the approximate integrals of motion, satisfy the 
condition \eqref{(2.18)}, so that  it results
\begin{equation}{\mathcal L}^ * G({\underline{\l}(\,\cdot\, ,t)})=O(\e).\label{(3.20)}
\end{equation}
So, in a local Gibbs state, in sufficiently small region and for a sufficiently small time the state looks like being in equilibrium, hence the name  {\it local equilibrium}. This situation is particularly convenient to describe a regime close to the hydrodynamics and, as discussed in \cite{M}, \cite{De M}, gives formally the Euler equations. However, this is not sufficient to  describe the Navier-Stokes regime and the ghost regime we are going to discuss here.

To show how the conservation laws give the hydrodynamic equations 
 we follow a procedure similar to the Hilbert expansion
proposed to  approximate the solutions of the Boltzmann equation.
Let us start with the phase space distribution function $$F_\e( x_1,\dots ,x_N, v_1,\dots, v_N, t)=F(\e^{-1}x_1,\dots, \e^{-1}x_N, v_1,\dots, v_N,\e^{-2}t)$$ for
the rescaled  system, which satisfies the rescaled Liouville equation
\begin{equation} {\partial F_\e \over \partial t} =\e^{-2}\mathcal{L}^ *  F_\e.
 \label{(2.25)}\end{equation}

We prepare the system at time zero as described by a local equilibrium distribution $\bar G_0$ defined as 
\begin{align}\bar G_0 &=Z_0 ^{-1}\exp \{ \{ \sum_{i=1}^N\sum_{\mu=0}^{d+1} \lambda_0 ^\mu(x_i,0) z_i ^\mu
\},\\
 \lambda _0^\mu (x,0) & =0, \mu=1,\dots ,d;\quad \la_0^0(x,0)\ne 0 ,\la_0^{d+1}(x,0)\ne 0, 
 \label{(2.27)}\end{align} 
 $\lambda_0^0$ and $\lambda_0^{d+1}$ functions  of $(x,t) $  with $x$-gradient of order $1$. Note that $\bar G_0 $ is not a stationary state for \eqref{(2.25)}.

Writing $ F_\e$ as a series in $\e$, $ F_\e=\sum_n \e^n F^n$, and
substituting it 
in \eqref{(2.25)} we get the diverging terms $\e^{-2}\mathcal{L} ^ *
F_0$ and $\e^{-1}\mathcal{L} ^ *
F_1$ which we have to take care of.
 
 Due to the initial conditions $\l_0^\beta(x,0)=0$,  for $\beta=1, \dots, d$, in absence of external forces we have $\l_0^\beta(x,t)=0$ at the lowest order, and we are forced  to choose  $F_0$ as  a local
Gibbs state $G_0$ with $\la_0^0$ and  $\la_0^4$ functions of $(x,t)$ to be determined while $\la_0^\beta=0$, for $\beta=1,\dots,d$. 

To single out the non-hydrodynamic contribution 
to $F_\e$ let us decompose $F_\e$ in a part
which is Gibbsian with parameters slowly depending on the
microscopic variables and depending on $\e$ by means of a series
in $\e$, and a remainder. More explicitly, we put 
\begin{equation}F_\e =G_\e + \e  G_0R_\e , \label{(2.26)}\end{equation}
with
\begin{equation*}G_\e =Z_\e ^{-1}\exp \{ \sum_{i=1}^N\sum_{\mu=0}^{d+1} \lambda_\e ^\mu(x_i,t) z_i ^\mu
\},\end{equation*}
\begin{equation} \lambda_\e ^\mu (x,t) =\sum _{n=0} ^{\infty} 
\e^n \lambda^\mu _n (x,t);\phantom{...} \lambda^\mu _0 =0, \mu=1,\dots ,d;\quad \la_0^0(x,t)\ne 0 ,\la_0^{d+1}(x,t)\ne 0, 
 \label{(2.27.1)}\end{equation}
 where $\lambda_n^\mu$  are functions  of $(x,t)$  to be determined. Note that, if $\lambda_0^0$ and $\lambda_0^{d+1}$ do not depend on $(x,t)$, we are back to to the case of the incompressible Navier-Stokes-Fourier system discussed in \cite{EM}.

In our context $G_0$, the zero order term in the expansion, 
is a particular  time dependent {\it local} equilibrium, but not all the relevant terms are included in $G_0$. We include all the
hydrodynamic terms in $G_\e$  and we can assume that in $R_\e$ there
are no terms which are combinations  of the invariant
quantities $z^\alpha$ with coefficients depending on the macroscopic
variables, since  these terms are already present in $G_\e$. $R_\e$ represents the non-equilibrium part of the 
distribution $F_\e$, which takes into account the fast modes in
the system, namely the non-conserved quantities. 
They appear at the hydrodynamic level only through dissipative 
effects  and determine the expression of transport 
coefficients.

To make this concept more precise we refer to 
\cite {S1}, \cite {S} where it is introduced
the Hilbert space of the local observables
equipped with the scalar product
\begin{equation}(\phi, \psi)=\int dx [\langle \phi\tau_x \psi\rangle_{G_0} -
\langle \phi\rangle_{G_0} \langle \psi\rangle_{G_0}]. \label{(2.23)}\end{equation}
Here $\langle \cdot \rangle_{G_0}$ is the average on the local Gibbs 
measure $G_0$.  In terms of this scalar product we define the projector 
on the invariant space  as
\begin{equation}{\mathcal{P} \phi}= \sum_{\mu,\nu=0}^{d+1} (\phi,z^\mu) (z,z)^{-1}_{\mu \nu}
 z^\nu, \label{(2.24)}\end{equation}
where $(z,z)^{-1}$ denotes the inverse of the matrix with elements
$\langle z_\mu z_\nu\rangle_{G_0}$. 

We assume that $R_\e$ has no component on the invariant space, in other words we ask
\begin{equation}{\mathcal P}[R_\e ]=0. \label{(2.28)}\end{equation}
\noindent We also assume that 
\begin{equation} G_0R_\e(t) =G_0R_1+\e G_0R_2 (t) +O(\e^2) .  \label{(2.29)}\end{equation}

We need  explicit expressions for $R_i,i=1,2$, in terms of the empirical 
fields, so that, inserting \eqref{(2.26)} in the conservation laws
averaged with respect to $F_\e$, we can get close equations for the
empirical fields up to order $\e$. To find such  expressions, 
we insert the expansion \eqref{(2.26)} for $F_\e$ in the Liouville 
equation \eqref{(2.25)} 
\begin{align}\begin{split}
\partial_t G_\e + \e \partial_t(G_0R_\e)= \e^{-2}\mathcal{L}^ *  G_\e +\e^{-1} \mathcal{L}^ * G_0
R_\e \\
=\e^{-2}\mathcal{L}^ *  G_0+\e^{-1}\mathcal{L}^ * G_1+\e^{-1} \mathcal{L}^ * G_0
R_1+\mathcal{L}^ * G_0R_2 +O(\e) .
\label{(2.30)}
\end{split}\end{align}  
By \eqref{(3.20)}, $\e^{-1}\mathcal{L}^ * G_0=O(1)$. 
In fact,
 the expression of $\mathcal{L}^ * G_0$  can be computed as
\begin{equation}
\mathcal{L}^ * G_0=\e G_0\sum_{i=1}^N \sum_{\mu=0,d+1}\sum_{\ga=1}^d
{\partial_\gamma \lambda_0 ^\mu 
} (x_i) w^{\mu  \gamma} _i =\e G_0\sum_{i=1}^N [{\partial_\gamma \lambda_0 ^0 
} (x_i) w^{0  \gamma} _i+{\partial_\gamma \lambda_0 ^{d+1}
} (x_i) w^{{d+1}  \gamma} _i].\label{(2.33.1)}
\end{equation}
Next we compute  $\mathcal{L}^ * G_1$. We write
\begin{equation}G_1 = G_0 g_1, 
 \label{(2.32)}\end{equation}
where 
\begin{equation} g_1 =  \sum_{j=1}^N\sum_{\mu=0}^{d+1} \la_1^\mu(x_j,t) [z_j ^\mu - <z_j ^\mu>_{G_0}].\label{g1}\end{equation}
Since $g_1$ is a linear combination of the 
invariant quantities $z$ with
coefficients depending on the macroscopic variables, the action of
$\mathcal{L}^ *$ on it  gives a linear combination of the currents
$w$ with a factor $\e$:    
\begin{equation}
-\e^{-1}\mathcal{L}^ * G_1= \sum_{i=1}^N \sum_{\mu=0}^{d+1}\sum_{\ga=1}^dG_0
{\partial_\gamma \lambda_1 ^\mu\
} (x_i,t) w^{\mu  \gamma} _i + \e^{-1}g_1 \mathcal{L}^ * G_0= O(1) .
\end{equation}

We conclude that  $\e^{-1}\mathcal{L}^ * G_1$, $\partial_t G_\e $ and  $\mathcal{L}^ * G_0
R_2$  are of order at least $1$.

If we multiply by  $\e$ \eqref{(2.30)},
 in the limit $\e\to 0$, we have
\begin{align} \e^{-1}\mathcal{L}^ *  G_0+ \mathcal{L}^ * G_0
R_1=0 \label{(2.31.2)}.\end{align} 

Moreover, it is easy to check that ${\mathcal{L}}^ * G_0$   is odd by the exchange $v\to -v$ because $w^{\mu  \gamma} _i$, for $\mu=0,d+1\quad \gamma=1\cdots d$ is odd in $v$. Hence,    the  condition \eqref{(2.31.2)} can determine only the odd part of $R_1$ that we call $R_1^a$.  

To find $R_1^s$, the even part of $R_1$,  we apply again ${\mathcal L}^*$ to \eqref{(2.31.2)} to get, in the limit $\e\to 0$,
\begin{equation}
\mathcal{L}^ *  \mathcal{L}^ * 
G_0R_1^s=-\e^{-1} \mathcal{L}^ *  \mathcal{L}^ * G_0.
\label{(2.31.1)}\end{equation}
The term 
 $\e^{-1}\mathcal{L}^ *  \mathcal{L}^ * G_0 $ is even (see Appendix A.1)  so that this determines  $R_1^s$. It has a part of order $\e$ which  enter in the form of the new transport coefficients in the momentum equation and a part 
 of order $1$ which does not give contribution in the equation.  The expression of  $\e^{-2}\mathcal{L}^ *  \mathcal{L}^ * G_0 $ is very complicate and involves two space derivatives of the chemical potentials. 
To summarize, $R_1=R_1^a+R_1^s$ with $R_1^a$ odd of order $1$ and $R_1^s$ even  solutions of \eqref{(2.31.2)} and \eqref{(2.31.1)}.

Now we go back to \eqref{(2.30)} and get, in the limit $\e\to 0$,
\begin{equation}
{\mathcal L}^*G_0R_2=-\e^{-1}{\mathcal L}^*G_1-\partial_t G_0.\label{(2.31)}\end{equation}

\noindent This condition determines $R_2$ in the limit $\e\to 0$ as
\begin{equation}{\mathcal{L}^ *   G_0R_2}=G_0\sum_{i=1}^N \sum_{\mu=0}^{d+1}\sum_{\ga=1}^d
{\partial_\gamma \lambda_1 ^\mu
} (x_i,s) w^{\mu  \gamma} _i -\e^{-1}g_1\mathcal{L}^ * G_0-\partial_t G_0.\label{(2.33)}
\end{equation}

In order the previous equations to be satisfied the {r.h.s.} cannot  have components on the null space. This condition will be satisfied a-posteriori using the fact that $(\rho,e,u)$ is solution of the hydrodynamical equations (in Appendix A.2).

We assume that there exists a unique
solution $R_1^a(t)$ to \eqref{(2.31.2)} and $R_2$ to \eqref{(2.33)} such that ${\mathcal P}R_1^a(t)=0$, ${\mathcal P}R_2(t)=0$, which are expressed formally in terms of $
{\mathcal{L}^ *}^{-1} $. We assume also that there exists a unique solution to \eqref{(2.31.1)} with ${\mathcal P}R_1^s(t)=0$ which is expressed formally in terms of ${\mathcal{L}^ *}^{-1}{\mathcal{L}^ *}^{-1}$.
This is the assumption we really need on the 
inverse of $\mathcal{L}^ *$ to get the result.

For sake of simplicity, we consider from now on a particular form for $g_1$, namely we put to zero $\lambda_1^0$ and $\lambda_1^{d+1}$:
\begin{equation} \lambda_1^0=0,\quad \lambda_1^{d+1}=0. \label{L10}\end{equation}
This is sufficient to obtain the limiting equation for $(\rho, e, u)$. Assuming a general form of $g_1$ we would also get equations for the first corrections $\rho_1$ and $e_1$, which are not needed to obtain the ghost equations \eqref{fluid system-1}.

\vskip .5cm

\section{Hydrodynamic equations}

\vskip.5cm
The incompressible limit corresponds to the assumption 
that the velocity field is small compared with the sound speed. 
In other words we assume that $U^\mu(x,t)\equiv\langle
z^\mu(x)\rangle_{F_\e(t)}$, $\mu=1,\dots,d$, starts with a term 
of order $\e$. Under the assumptions on $F_\e$, this corresponds to
choose  $\la_0^\mu=0$ for $\mu=1,\dots,d$.  Moreover, it
results $U^\mu(x,t)=\e \rho T\la_1^\mu(x,t)+O(\e^2)$ 
with $T$, the  temperature of the Gibbs state $G_0$,
given by $(\lambda_0^{d+1})^{-1}$ and $\rho$ the density of the
Gibbs state $G_0$ corresponding to the   chemical
potential $\la_0^0$. We denote by $u^\mu(x,t)$ the rescaled velocity
field given by $u^\mu(x,t)=T\la_1^\mu(x,t)$. For sake of simplicity, we will not write the explicit dependence on time in the chemical potentials from now on.

We will use the notation $\Big\langle f \Big\rangle_{G_0}=\Big\langle G_0f\Big\rangle$ where $\Big\langle \cdot\Big\rangle$ is the  already introduced integration  w.r.t. the {\it a priori} measure $d\mathbb{Z}={1\over n!}d^d x_1 d^d v_1$  $\dots d^d
 x_N d^d v_N$.

\vskip.1cm

\subsection{\bf Continuity equation}
\vskip.1cm

Let's derive first
  the continuity  equation. To obtain it, we
start from the conservation law for the empirical  density \eqref{(2.7)}
and we take the expectation with respect to the  non-equilibrium
measure $F_\e(t)$ 
\begin{equation}\Big\langle \e^d \sum_{i=1}^N f(x_i) \Big\rangle _{F_\e (t)} - 
\Big\langle \e^d \sum_{i=1}^N f(x_i) \Big\rangle _{F_\e (0)} = 
\e^{-1} \int_0 ^t ds 
\Big\langle \e^d \sum_{i=1}^N\sum_{k=1}^d{\partial_k f}(x_i)
v_i^k \Big\rangle _{F_\e (s)} .
 \label{(3.1)}\end{equation}
Using  \eqref{(2.29)} and the fact that $\Big\langle \e^d \sum_{i=1}^N f(x_i) \Big\rangle _{G_0} =\int dx f(x)\rho(x,t)$ we see that the l.h.s. of 
\eqref{(3.1)} gives 
$$
\int dx f(x)\rho(x,t)-
\int dx f(x)\rho(x,0)+O(\e).
$$
The term
$$\e^{-1} \int_0 ^t ds 
\Big\langle \e^d \sum_{i=1}^N\sum_{k=1}^d{\partial_k f}(x_i)
v_i^k \Big\rangle _{G_0}, $$
in r.h.s. of \eqref{(3.1)} does not appear since
 $G_0$ is Gaussian in the velocities with zero mean and the rhs of \eqref{(3.1)} becomes
\begin{equation}\int_0 ^t ds \Big\langle g_1 \e^d\sum_{i=1}^N \sum_{k=1}^d{\partial_k
f} (x_i) v_i^k  \Big\rangle_{G_0}  + \int_0 ^t ds \Big\langle \e^d \sum_{i=1}^N \sum_{k=1}^d{\partial_k
f} (x_i) v_i^k  R_1\Big\rangle_{G_0} +O(\e).\label{(3.1.1}
\end{equation}
The second term in \eqref{(3.1.1} is $0$ by using that ${\mathcal P}R_1=0$. Now we discuss the first term. By \eqref{g1},
\begin{equation}\int_0 ^t ds \Big\langle g_1 \sum_{i=1}^N\sum_{k=1}^d{\partial_k
f} (x_i) v_i^k  \Big\rangle_{G_0}   =
\int_0 ^t ds \Big\langle \sum_{j=1}^N \sum_{\mu=1}^{d}\lambda^\mu
_1(x_j)  z^\mu_j  \e^d \sum_{i=1}^N\sum_{k=1}^d{\partial_k f
}(x_i) v_i^k \Big\rangle_{G_0} . \label{(3.2)}\end{equation}

We use  the choice we made for $\lambda_1$ so that the l.h.s. of \eqref{(3.2)} becomes
\begin{equation}\Big\langle \e^d \sum_{i=1}^N\sum_{\mu,k=1}^d \lambda^\mu _1(x_i) 
{\partial_\mu f}(x_i) v_i^\mu v_i^k \Big\rangle_{G_0}. 
 \label{(3.3)}\end{equation} %
Since the average on $G_0$ of $v_i^\mu v_i^k$ contributes only
for $k=\mu$, in the limit $\e \to 0$ we have,  using $u^\mu(x,t)=T\la_1^\mu(x,t)$,
\begin{equation}-\int_0 ^t ds \int dx \sum_{\mu=1}^d {\partial_\mu (\rho\/u^\mu )
}(x,t) f(x), \label{(3.4)}\end{equation}
for any test function $f$ and for any $t$.
Hence
\begin{equation}\rho_0(x,t)-
\rho(x,0)=-\int_0^t \text{div}\/(\rho u) , \label{(3.5)}\end{equation}
or in differential form 
$$\partial_t\rho=-\text{div}\/(\rho u).$$

\vskip.1cm

\subsection{\bf Pressure}
\vskip.1cm
We examine now the second conservation law \eqref{(2.8)}. By averaging 
as before, for $\be=1, \dots, d$, we get, in the limit $\e\to 0$, 
\begin{align}\begin{split}& 
\Big\langle \e^d \sum_{i=1}^N f(x_i) v_i ^\be 
\Big\rangle_{F_\e (t)} -  \Big\langle \e^d \sum_{i=1}^N f(x_i)  v_i
^\be \Big\rangle_{F_\e (0)} \\&=\e^{-1} \int_0 ^t ds 
\Big\langle  \e^d \int dx\sum_{k=1}^d 
{\partial_k f}(x)w^{\beta k}(x)\Big\rangle_{F_\e} + 
O(\e) .\label{(3.6)} 
\end{split}\end{align} 
Using the assumptions on $F_\e(t)$ we see that 
the l.h.s. is  of order $\e$, 
since the term of order $1$ vanishes because
$\lambda_0^\mu,\mu=1,2,3$ are zero.

We write the current $w^{\be k}$ using the decomposition  \eqref{(2.145)} as $w^{\be k}_*+\e{\partial_\ga 
 } \bar\Phi_0^{k\beta\gamma }(x)+O(\e^2)$,  with $w_*^{\beta k}$ given by \eqref{w*}.
Using the assumptions \eqref{(2.26)} and \eqref{(2.29)} we have:
\begin{equation}\e^{-1} \Big\langle w_*^{\be k} \Big\rangle _{F_\e}= 
\e^{-1}\Big\langle w_*^{\be k} \Big\rangle_{G_\e}
+  \Big\langle w_*^{\a\be} R_\e \Big\rangle,\label{(3.7)}\end{equation}
so  that
\begin{equation}\e^{-1} \Big\langle w^{\be k} \Big\rangle _{F_\e}= 
\e^{-1}\Big\langle w^{\be k}_*\Big\rangle_{G_\e}+\Big\langle\sum_{\ga=1}^d{\partial_\ga 
 } \bar\Phi_0^{k\beta\gamma }(x)\Big\rangle_{G_0}
+  \Big\langle w_*^{\a\be} R_\e \Big\rangle_{G_0} + O(\e).\label{(3.7.1)}\end{equation}
The second term on the r.h.s. is zero since the term $\bar\Phi_0^{k\beta\gamma }(x)={1\over 2}\e^d \sum_{i,j=1}^N \delta(x_i -x)\Phi_0^{\beta\gamma\nu}(\e^{-1}|x_i -x_j| )$ is antisymmetric under the exchange $i\to j$ while $G_0$ is symmetric.

We introduce the currents $\tilde w^{\be k}$ as given by  $ w^{\be k}_*$ with the velocities $v_i$ replaced by 
$\tilde {v}_i= v_i -\e u(x_i)$.  Then 
\begin{equation} w_*^{ \be k}(x)  ={\tilde w}^{ \be k} (x)+ \e^2 u^k(x) u^\be(x)
+\e u^k(x) \tilde v ^\be+\e u^\be(x) \tilde v ^k.\label{(3.8)}\end{equation}
For the symmetry of the measure $G_\e$ we have 
$\Big\langle \tilde w^{ \be k} (x) \Big\rangle_{G_\e} =O(\e^2)${,
if $k \ne \be$. The average of $\tilde w^{\be \be}$, 
$\be=1, \dots, d$, with respect
the local Gibbs state $G_\e$ is, by the virial theorem, the
thermodynamic pressure $P_\e$ in the state  $G_\e$ \cite {P}.

Now we consider the term $ \Big\langle w_*^{\a\be} R_\e \Big\rangle_{G_0}$ and use 
\eqref{(2.33.1)}, \eqref{(2.31.2)} and 
the `identity' ${\mathcal L^*}^{-1}{\mathcal L^*} R_1^a=R_1^a$. For sake of shortness we omit the time dependence in the chemical potentials. We obtain \begin{equation}
\Big\langle w_*^{\a\be} R_\e \Big\rangle_{G_0}= \Big\langle \tilde w_*^{\a\be} R_1^a\Big\rangle_{G_0} + O(\e)
=\Big\langle  \sum_{i=1}^N \sum_{\ga=1}^d[{\partial_\ga \lambda_0 ^0 
} (x_i) w^{0  \gamma} _i+{\partial_\ga \lambda_0 ^{d+1} 
} (x_i) w^{{d+1}  \gamma} _i]{\mathcal L}^{-1}\tilde w^{\a\be}\Big\rangle_{G_0}+ O(\e).
 \end{equation}
 We have
 $$\int dx {\partial_k f}(x)\Big\langle  w_*^{\a\be} R_\e \Big\rangle_{G_0}=
 \int dx {\partial_k f}(x)\Big\langle  {\mathcal L}^{-1} w_*^{\a\be}\sum_{i=1}^N \sum_{\ga=1}^d[{\partial_\ga \lambda_0 ^0
} (x_i) w^{0  \gamma} _i+{\partial_\ga \lambda_0 ^{d+1}
} (x_i) w^{{d+1}  \gamma} _i]\Big\rangle_{G_0}+O(\e)
 $$
 $$
 = \e^{-d}\int dx {\partial_k f}(x)\int dy \sum_{\nu=0,d+1}\sum_{\ga=1}^d{\partial_\ga \lambda_0 ^\nu\
}(y) \Big\langle {\mathcal L}^{-1} w_*^{\a\be}(x)w^{\nu\gamma} (y)\Big\rangle_{G_0}+O(\e).
$$
The symmetries of the microscopic current-current correlations 
imply (see \cite {S}) that the cross correlations between $\mu=0,d+1$ and $\beta=1,2,3$ vanish. 
 
 Summarizing, 
 eq. \eqref{(3.6)}  implies  for $P^\e=\Big\langle w_*^{\beta\beta}\Big\rangle_{G_\e}$
\begin{equation}\e^{-1} \int dx \nabla  f(x) P^\e(x,t)
=O(\e) . \label{(3.9)}\end{equation}
Since $P^\e$ is a function of the thermodynamic parameters
$\lambda_\e$, we can expand it in series of $\e$ as $\sum_k \e^k
P_k$, where  $P_k ={ {d^k P^\e \over d \e^k} \big |_{\e=0}}$
We have  that $P_0$  is a 
function of the  $\la_0^0$ and $\lambda_0^{d+1}$, while 
$P_1  =\sum_{\mu=0}^{d+1}{\partial P^\e \over  \partial
\lambda_\e ^\mu} \big|_{\e=0}
\lambda_1 ^\mu $. 
In order to fulfill \eqref{(3.9)} for any test function
$f$, 
$$\nabla P_0=0,\quad \nabla P_1=0.$$ 
In particular, the pressure $P_0(\rho, T)$ is a function of $\rho, T$  quantities conjugate to $\la_0^0$ and $\lambda_0^{d+1}$ respectively, and has to be constant versus $x$. In the following, we will denote $P_0$ simply as $P$.

\vskip.2cm

\subsection{\bf Momentum  equation}

\vskip.2cm

To determine the equation for $u^\mu(x,t)$, which is of order $\e$, we have to rescale
the empirical velocity field. This means that
we have to look at the empirical field 
\begin{equation}\bar z^\a(x) = \e^{-1}\e^d \sum_{i} v^{\a} _i \delta (x_i -x),
\phantom{...}\a =1,\dots,d. \end{equation}
We proceed as we did before to obtain \eqref{(3.6)}, but we have to 
look at the explicit form of the terms $O(\e)$ and $O(\e^2)$ because they have to
be divided by $\e^2$. We have:
\begin{align} 
\begin{split}
&\Big\langle \e^{d-1} \sum_{i=1}^N f(x_i) v_i ^\be 
\Big\rangle_{F_\e (t)} -  \Big\langle \e^{d-1} \sum_{i=1}^N f(x_i)  v_i
^\be \Big\rangle_{F_\e (0)} =\\  & \e^{-2} \int_0 ^t ds
\Big\langle  \e^d \sum_{i=1}^N\sum_{k=1}^d  {\partial_k f
}(x_i)\{ v_i^k v_i^\be  +  {1 \over 2}\sum_{j \ne i}
\Psi^{\be k}(\e^{-1}(x_i -x_j))  \} \Big\rangle _{F_\e (s)} +\\
 &+ \e ^{-1}\int_0 ^t ds\Big\langle{1 \over 4} \e^d \sum_{i,j=1}^N\sum_{\ga,\nu=1}^d{\partial_{\ga\nu}^2 f
}(x_i) \Phi_0^{\nu\beta\gamma}(\e^{-1}(x_i  -x_j ))\Big\rangle _{F_\e (s)}\\&+ \int_0 ^t ds\Big\langle
{1 \over 12}\e^d \sum_{i,j=1}^N\sum_{\ga,\nu,\a=1}^d{\partial_{\ga\nu\a}^3 f
 }(x_i) \Phi^{\a\beta\gamma\nu}(\e^{-1}(x_i  -x_j ))\Big\rangle_{F_\e(s)} \\
  \label{(3.10)}
\end{split}
 \end{align}
The l.h.s. of \eqref{(3.10)} converges to
\begin{equation}\int dx f(x) \rho [u(x,t) - u(x,0)], \label{(3.11)}\end{equation}
as $\e\to 0$.

To get the equation for the velocity field we have to compute the non-equilibrium 
average of the velocity current tensor $w^{ \be k}$ but now there
is a factor $\e^{-2}$ in front of it.
Therefore we see that in this case also the terms of order 
$\e^{2} $ in  \eqref{(2.26)} have to be taken into account. 

We start by discussing the term
containing $D_0$   namely the second line in the r.h.s. of \eqref{(3.10)}. 
The lowest  order term in $\e$ is
$$
\e ^{-1}\int_0 ^t ds\Big\langle{1 \over 2} \e^d \sum_{i,j=1}^N\sum_{\ga,\nu=1}^d{\partial_{\ga\nu}^2 f
}(x_i) \Phi_0^{\beta\gamma\nu}(\e^{-1}(x_i  -x_j ))[G_0+\e G_0g_1+\e G_0R_1]\Big\rangle.
 $$
 
 The diverging term
$$
\e ^{-1}\int_0 ^t ds\Big\langle{1 \over 2} \e^d \sum_{i,j=1}^N\sum_{\ga,\nu=1}^d{\partial_{\ga,\nu}^2 f
}(x_i) \Phi_0^{\beta\gamma\nu}(\e^{-1}(x_i  -x_j ))\Big\rangle_{G_0}.
$$
is zero by symmetry property of the potential. In fact, $\Phi_0^{\beta\gamma\nu}$ is antysimmetric in the exchange $\xi_i\to\xi_j$ while $G_0$ is symmetric.

The term of order zero in $\e$ is 
$$
\int_0 ^t ds\Big\langle{1 \over 2} \e^d \sum_{i,j=1}^N\sum_{\ga,\nu=1}^d{\partial_{\ga\nu}^2 f
 }(x) \Phi_0^{\beta\gamma\nu}(\e^{-1}(x_i  -x_j ))[ G_0g_1+ G_0R_1^a]\Big \rangle .
 $$
 For the same reasons the part involving $g_1$ is zero because all the conserved quantities are symmetric. We are left with 
 $$
\int_0 ^t ds\Big\langle{1 \over 4} \e^d \sum_{i,j=1}^N\sum_{\ga,\nu=1}^d{\partial_{\ga\nu}^2 f
 }(x_i) \Phi_0^{\beta\gamma\nu}(\e^{-1}(x_i  -x_j ))R_1^a\Big\rangle_{G_0} .
 $$
We can safely change $ \Phi_0^{\beta\gamma\nu}$ in $\tilde \Phi_0^{\beta\gamma\nu}:=\Phi_0^{\beta\gamma\nu}-{\mathcal P}\Phi_0^{\beta\gamma\nu}$ since ${\mathcal P} R_1$ is zero.

 By \eqref{(2.28)} we can use the `identity'
${(\mathcal{L} ^ *})^{-1}{\mathcal{L} ^ *} R_1^a =R_1^a$
and the expression of $R_1^a$ to get
$$
\int_0 ^t ds\Big\langle{1 \over 4} \e^d \sum_{i,j=1}^N\sum_{\ga,\nu=1}^d{\partial_{\ga\nu}^2 f
 }(x_i)  {\mathcal L}^{-1} \tilde \Phi_0^{\beta\gamma\nu}(\e^{-1}(x_i  -x_j ))
\sum_{k=1}^d [{\partial_k \lambda_0 ^0
} (x,s) w^{0  k} _i+{\partial_k \lambda_0 ^{d+1}
} (x,s) w^{{d+1}  k} _i]\Big\rangle_{G_0}.
 $$

 This term is zero by oddness in the velocity.
 
We observe that the term containing   $D$, third line in \eqref{(3.10)} is  not  
$0$ because the lowest order is given by an average
with respect to $G_0$  (Gibbs measure with non constant parameters).  The
other terms are  of higher order, hence  do not contribute.

The explicit expression is 
\begin{align}
\begin{split}
&{1\over 6}\int dx \sum_{\a,\ga,\nu=1}^d{\partial_{\a\ga\nu}^3 f
}(x)\Big\langle \bar\Phi_{\beta\a\gamma\nu}(x)\Big\rangle _{F_\e}= \label{99}\\
&\Big\langle{1 \over 12}\e^d\sum_{i \ne j=1} ^N
\sum_{\a,\ga,\nu=1}^d{\partial_{\a\ga\nu}^3 f
 }(\e\xi_i)\partial_\be V (|\xi_i -\xi_j|) [\xi_i ^\ga 
-\xi_j^\ga ] [\xi_i ^\nu  -\xi_j ^\nu ] [\xi_i ^\a  -\xi_j^\a ]\Big\rangle _{G_0} + o(\e).
\end{split}\end{align}

By symmetry,  only the terms of the form $\Big\langle\bar\Phi_{\alpha \alpha\beta\beta}\Big\rangle _{G_0} $
are different from zero. Define $ \Phi_{ \alpha\beta}:=\bar\Phi_{\alpha \alpha\beta\beta}$ and $\hat\Phi_{ \alpha\beta}:=\Big\langle\Phi_{ \alpha\beta}\Big\rangle _{G_0}$.

\noindent $\hat\Phi_{ \alpha\beta}$ is a function of $\rho, T$ and using the condition on $\nabla P_0=0$ it can be seen as a function of $T$ only, so  only the gradient of $T$ will appear in the final term.  By two   integrations by parts we move two $x$-derivatives in the main term in \eqref{99} on $\hat\Phi_{ \alpha\beta}$ and \eqref{99} becomes 
$${1\over 6}\int dx  \sum_{\a=1}^d{\partial_\a f} {\partial_{\a\beta}^2 
} \hat\Phi_{\alpha \beta}(\rho, T).
 $$

We have 
$$
 {\partial_\a } \hat\Phi_{ \alpha\beta}={\partial\hat\Phi_{ \alpha\beta}\over \partial T} {\partial_\a }T:=\hat\Phi'_{ \alpha\beta}{\partial_\a }T,
$$
and
$$
\partial^2_{\a\beta}\hat\Phi_{ \alpha\beta}=\partial_\beta(\hat\Phi'_{ \alpha\beta}{\partial_\a }T)
={\partial_\a }T\partial_\beta(\hat\Phi'_{ \alpha\beta})+\hat\Phi'_{ \alpha\beta}\partial^2_{\a\beta}T$$
$$={\partial_\a }T{\partial \hat\Phi'_{ \alpha\beta}\over \partial T}\partial_\beta T+\hat\Phi'_{ \alpha\beta}\partial^2_{\a\beta}T
=\hat\Phi''_{ \alpha\beta}{\partial_\a }T\partial_\beta T+\hat\Phi'_{ \alpha\beta}\partial^2_{\a\beta}T.
$$
Therefore
\begin{align}\begin{split}&{1\over 6}\int dx  \sum_{\a=1}^d{\partial_\a f} {\partial_{\a\beta}^2 
 } \hat\Phi_{\alpha \beta}(\rho, T)=\\
 & \sum_{\a=1}^d \Big[\int dx  \partial_\beta f [Y_1({\partial_\a }T\partial_\beta T-\delta_{\a\beta}{1\over d}\sum_{\gamma=1}^d({\partial_\gamma }T)^2) +{\bar\omega_1}  \delta_{\a\beta}\sum_{\gamma=1}^d({\partial_\gamma }T)^2\\&+ Y_2 (\partial^2_{\a\beta}T-\delta_{\a\beta}{1\over d}\sum_{\gamma=1}^d\partial^2_{\gamma\gamma}T)+{\bar\omega_2}\delta_{\a\beta}\sum_{\gamma=1}^d\partial^2_{\a\gamma}T)]\Big],
 \label{Y}
\end{split}\end{align}
 where 
 
\begin{equation}Y_1={1\over 6}\hat\Phi_{ \alpha\beta}'',\quad \alpha\ne \beta; \quad Y_2={1\over 6}\hat\Phi'_{ \alpha\beta}\quad \alpha\ne \beta; \quad\bar\omega_1={1\over 6d}\sum_{\gamma=1}^d\hat\Phi_{ \gamma\gamma}'',\quad \bar\omega_2={1\over 6d}\sum_{\gamma=1}^d\hat\Phi'_{ \gamma\gamma}.
\label{YY}
\end{equation}

Finally, we turn to first term in the rhs of \eqref{(3.10)} 
and   compute the contributions for the currents $w_*$ and use the definition \eqref{(3.8)} of $\tilde w$:

 \begin{align} 
 \begin{split}
 &\e^{-2}\Big\langle w_*^{ \be k} (x) \Big\rangle_{G_\e} 
+ \Big\langle w_*^{ \be k} (x) R_\e\Big\rangle_{G_0} =
\e^{-2}\Big\langle \tilde w ^{\beta k}\Big\rangle_{G_\e}
+\rho u^\beta u^k + \e^{-1}\Big\langle  w_* ^{\beta k} R_\e
\Big\rangle_{G_0} +O(\e)=\\ & \e^{-2}P_0 +\e^{-1}P_1 + P_2 +\rho u^\beta
u^k +\e^{-1}\Big\langle{G_0R_1}  
w_*^{ \be k} (x) \Big\rangle +\Big\langle{G_0R_2}  
w_*^{ \be k} (x) \Big\rangle  +O(\e) .
\end{split}\label{(3.12)}\end{align}
The first two terms in the second line of \eqref{(3.12)}  do not  contribute
 because $P_0$ and $P_1$ are constant.
The fourth term in the rhs of \eqref{(3.12)}  gives the non linear transport term,
while $P_2$ represents the second order correction to the
thermodynamic pressure $P_\e$ and gives rise to
the unknown pressure $\frak p$ appearing in equation \eqref{fluid system-1}.

Now we pass to examine the terms involving $R_i$. To compute these terms, let us first introduce
$\bar w^{ \be k}=w_*^{ \be k} - \mathcal{P}  w_*^{ \be k}$ 
and notice that \eqref{(2.28)} implies
\begin{equation}\Big\langle{R_1}\mathcal{P} w^{\a\be} (x) \Big\rangle_{G_0} =0,\quad  \Big\langle{R_2}\mathcal{P} w^{\a\be} (x) \Big\rangle_{G_0}=0. \label{(3.14)}\end{equation}

By \eqref{(2.28)} we can use the `identity'
${(\mathcal{L} ^ *})^{-1}{\mathcal{L} ^ *} G_0R_i =G_0R_i,\quad i=1,2$
and  \eqref{(2.33)}. 
First of all, let's consider the   diverging term due to $R_1$ and start from the contribution due to $R_1^a$: 

$$\e^{-1}\Big\langle\sum_{j =1}^N[{\partial_\ga \lambda_0 ^0 
} (x)  w^{0  \gamma} _j+{\partial_\ga \lambda_0 ^4
} (x,s) w^{4  \gamma} _j]\e^d \sum_{i=1}^N {\partial_k f }(x_i){\mathcal{L} } ^{-1}\bar w^{ \be k}_i
\Big\rangle_{G_0}.$$

The symmetries of the microscopic current-current correlations 
imply (see \cite {S}) that the cross correlations between $\mu=0,d+1$ and $\beta=1,2,3$ vanish. Hence this term is zero.

\noindent On the other hand the term involving $\e^{-1}R_1^s$, 
\begin{equation}\e^{-1}\Big\langle R_1^s\e^d \sum_{i =1}^N\bar w^{ \be k}_i{\partial_k f (x_i)}\Big\rangle_{G_0},
\label{RS}
\end{equation}
by our assumption is of order $1$ and has to be computed. 

This term gives in the hydrodynamic equations a term similar to the one due to $D$ and will add new terms to  the transport coefficients $Y_1$ and $Y_2$. The explicit expression is computed in the Appendix A.1. The total transport coefficients in the equation will be denoted by $K_1$ and $K_2$.

We discuss now the last term in the r.h.s of \eqref{(3.12)} involving $R_2$, 
\begin{align} 
\begin{split}&\int dx \sum_{k=1}^d{\partial_k f } 
\Big\langle\bar{w}^{ \be k}(x) (\mathcal{L} ^ *)^{-1}{\mathcal{L}
^ *}G_0R_2\Big\rangle=\Big\langle {\mathcal{L}
^ *} G_0R_2 {\mathcal{L} } ^{-1} \e^d \sum_{i=1}^N\sum_{k=1}^d\bar{w}^{\be k}_i
{\partial_k f }(x_i) \Big\rangle \nonumber \\
=& -\e^{-1}\Big\langle \Big[G_0{\mathcal{L}
^ *}g_1+ g_1 {\mathcal{L}
^ *} G_0-\e\partial_t G_0\Big]{\mathcal{L} } ^{-1} \e^d \sum_{i=1}^N\sum_{k=1}^d\bar{w}^{\be k}_i
{\partial_k f }(x_i) \Big\rangle .
\end{split}
\end{align}
The term $\p_t G_0{\mathcal{L} } ^{-1}\dots $ in the r.h.s. does not contribute because $\mathcal{L}^{-1}$ is orthogonal to $\p_t G_0$.

 \noindent The second term is 
\begin{align}
\begin{split}
\Big\langle g_1&\sum_{j=1}^N \sum_{\ga=1} ^d [{\partial_\ga \lambda_0 ^0
} (x) w^{0  \gamma} _j+{\partial_\ga \lambda_0 ^{d+1}
} (x) w^{{d+1}  \gamma} _j]{\mathcal{L} } ^{-1}\e^d \sum_{i=1}^N\sum_{k=1}^d \bar w^{ \be k}_i
{\partial_k f }(x_i)\Big\rangle_{G_0}\\
=&\sum_{\ell=1}^N\sum_{\mu,\ga=1} ^d \sum_{j=1}^N {\partial_\ga \lambda_0 ^0
} (x) \Big\langle \la_1^\mu(x_\ell) z_\ell ^\mu w^{0  \gamma} _j{\mathcal{L} } ^{-1}\e^d \sum_{i=1}^N\sum_{k=1}^d \bar w^{ \be k}_i
{\partial_k f }(x_i)\Big\rangle_{G_0}\\
&+\sum_{\ell=1}^N\sum_{\mu,\ga=1} ^d   \sum_{j=1}^N \la_1^\mu(x_\ell){\partial_\ga \lambda_0 ^{d+1}
} (x) \Big\langle z_\ell ^\mu w^{{d+1} \gamma} _j{\mathcal{L} } ^{-1}\e^d \sum_{i=1}^N \sum_{k=1}^d\bar w^{ \be k}_i
{\partial_k f }(x_i)\Big\rangle_{G_0}.\label{(GI)}
\end{split}
\end{align}
In \eqref{(3.18.1)} will appear a term of a similar form. All the terms together   (\eqref{(GI)}+second term in \eqref{(3.18.1)}) give in the hydrodynamic equation a term of the form
$$\sum_{j,k,l=1}^d\partial_j[\a_{\beta jkl}(T)u_k \partial_\ell T],$$
which is Galileian invariant for any temperature only if $\a_{\beta jkl}=0$ (see \cite{Bo} page 1071). So, $\a_{\beta jkl}$ has to be zero based on this physical consideration. In the Appendix A.3 we will prove that this is indeed the case.

\noindent We are left with the first term
\begin{align}\begin{split}
 \Big\langle \sum_{j=1}^N \sum_{\mu=1}^{d}\sum_{l, k =1}^d
{\partial_l\lambda _1 ^\mu } (x_j)
\bar w^{\mu l}_j {\mathcal{L} } ^{-1}\e^d \sum_i \bar w^{ \be k}_i
{\partial_k f }(x_i)\Big\rangle_{G_0}  .\nonumber\\
\end{split}
\label{main1}\end{align}

\noindent We remark that ${\mathcal{L} }^{-1}$ is ``well defined'' on $\bar{w}$ 
by the assumptions discussed in section 2. The substitution of the current $w$ with $\bar w$ is correct
because 
the range of  $\mathcal{L}^{-1}$ is orthogonal to $\mathcal{P} w$.

\noindent Since $\lambda_1 ^\mu=\displaystyle{{u^\mu\over T}}$ and $T$ are not costant, we have two different contribution to \eqref{main1}

\begin{align}\begin{split}
\sum_{j=1}^N \sum_{\mu=1}^{d}\sum_{l, k =1}^d\Big\langle
{\partial_l \lambda _1 ^\mu } (x_j)
 \bar w^{\mu l}_j {\mathcal{L} } ^{-1}\e^d \sum_{i=1}^N \bar w^{ \be k}_i
{\partial_k f }(x_i)\Big\rangle_{G_0}  \nonumber\\
\end{split}
\end{align}
\begin{equation}
=\sum_{j=1}^N\sum_{\mu=1}^{d} \sum_{l, k =1}^d\Big\langle
[{1\over T}{\partial_l u ^\mu } (x_j)-u ^\mu {1\over T^2}{\partial_l T } (x_j)]
 \bar w^{\mu l}_j {\mathcal{L} } ^{-1}\e^d \sum_{i=1}^N \bar w^{ \be k}_i
{\partial_k f }(x_i)\Big\rangle_{G_0} . \label{(3.18.1)}\end{equation}

The second term will combine with the terms in \eqref{(GI)} to eliminate the term in the equation which is not Galileian invariant (see Appendix A.3).

We discuss here the first term in \eqref{(3.18.1)}. To find the expression of the transport coefficients we consider
\begin{eqnarray} 
\begin{split}\Big\langle \sum_{j=1}^2\sum_{\mu,l=1}^{d} 
 &{1\over T}{\partial_l u ^\mu }
(x_j)
w^{\mu l}_j {\mathcal{L} } ^{-1}\e^d \sum_{i=1}^N \sum_{k=1}^{d} \bar w^{ \be k}_i
{\partial_k f }(x_i)\Big\rangle_{G_0}=\notag\\
&\sum_{\mu,k,l=1}^{d} \int dy {1\over T}{\partial_l u ^\mu } (y)\int 
\e^{-d} dz{\partial_k f }(z)
\Big\langle   \bar{w}^{\mu l}(y) {\mathcal{L} } ^{-1} \bar w^{ \be
k}(z) \Big\rangle_{G_0} .\label{(3.16)}
\end{split} \end{eqnarray}
Since the Gibbsian state $G_0$ is invariant under translations
on $\Bbb R ^d$
we have
\begin{equation}\text{r.h.s. of \eqref{(3.16)}}=\sum_{\mu,k,l=1}^{d}\int dy 
{1\over T}{\partial_l u ^\mu } (y) \int d \xi
{\partial_k f }(y + \e \xi)
\Big\langle  \bar{w}^{\mu l}(0) {\mathcal{L} } ^{-1} \bar{w}^{ \be k}
(\xi)\Big\rangle_{G_0} , \label{(3.17)} \end{equation}
where we have changed the variable $z$ in $\xi=\e^{-1}(z-y)$ 
absorbing  the factor $\e^{-d}$. 
Hence,  provided that $\langle  \bar{w}^{\mu l}(0) 
{\mathcal{L} } ^{-1}  \bar{w}^{ \be k}(\xi)  \rangle_{G_0}$ decay fast enough
for large $\xi$, up to $O(\e)$ we have  
\begin{equation}\text{r.h.s. of \eqref{(3.16)}})= \sum_{\mu,k,l=1}^{d}\int dy 
{1\over T}{\partial_l u ^\mu }(y)
{\partial_k f }(y) \int  d\xi 
\Big\langle  \bar{w}^{\mu l}(0) {\mathcal{L} } ^{-1}  \bar{w}^{ \be
k}(\xi)  \Big\rangle_{G_0}. \label{(3.18)}\end{equation}

\noindent To conclude the argument we transform  ${\mathcal{L}^{-1}}$ in a
time-integral of $\exp [t \mathcal{L}]$:
\begin{equation}-\Big\langle   \bar{w}^{\mu
l}(0) {\mathcal{L} } ^{-1} 
 \bar{w}^{ \be k}(\xi) \Big\rangle_{G_0}=
 \int_0 ^{\infty} d\tau 
\Big\langle   \bar{w}^{\mu l}(\xi, \tau)   \bar{w}^{ \be k}(0,0)
 \Big\rangle_{G_0}  .\label{(3.14)} \end{equation}
In this formula with an abuse of notation we are using the  symbol $\bar{w}^{\mu}(\xi)$ for the microscopic current. We notice that before the change of variable to microscopic variables $\xi$ the meaning of the current $\bar{w}^{\mu}$ was indeed $\bar{w}^{\mu}_\e(x)= \bar{w}^{\mu}(\e^{-1} x)$. 
We have that  (see \cite {S}) 
\begin{equation} \int  d\xi \Big\langle \bar{w}^{\mu l}(\xi, \tau) 
\bar{w}^{\be k}(0,0) \Big\rangle_{G_0} = 
c(\tau)[\delta_{k l}\delta_{\be k}
+ \delta_{k\mu} \delta_{\be l}] + c'(\tau)\delta_{ \be k}
\delta_{l \mu} .\label{(3.19)}\end{equation}
Therefore the time integral of \eqref{(3.19)} has only two independent 
coefficients
\begin{equation}\int_0^\infty d\tau c(\tau)=2 \eta T;\phantom{...} 
\int_0^\infty d\tau c'(\tau)=2T(\z -{2 \over d}\eta ),
 \label{(3.20bis)}\end{equation}
where $\eta$ and $\z$ are the shear viscosity and the bulk 
viscosity respectively.

\noindent They are finite if the correlations decay sufficiently fast to 
make the time integrals
in \eqref{(3.20bis)} convergent. Notice that the subtraction of $\mathcal{P}
\tilde w^{\a\be}$ has been crucial, 
because the self-correlation of the slow part of the current
does not decay in time.

\noindent Since  $\text{div} \ u\ne 0$, the term 
proportional to the bulk viscosity does appear in the limiting
equation. The computation gives also the Green-Kubo formula for the bulk 
viscosity $\zeta$
\begin{equation}\zeta ={1 \over 2 d^2 T} \int_0 ^{\infty} d\tau \int d\xi
\Big[\Big\langle \sum_\a \bar w ^{\a \a} (\xi,\tau) \sum_\ga 
\bar w ^{\ga \ga} (0,0)\Big\rangle_{G_0} - \Big\langle \sum_\a 
\bar w ^{\a \a}
\Big\rangle_{G_0} \Big\langle \sum_\ga \bar w ^{\ga \ga}
\Big\rangle_{G_0}\Big]. \label{(3.23)}\end{equation}
The usual expression given in \cite{S} is recovered using the
explicit form of the projector $\mathcal{P}$.
\vskip.1cm
The viscosity is given by
\begin{equation}\eta ={1 \over 2T} \int_0 ^{\infty} d\tau \int d\xi
\Big\langle \bar w ^{12} (\xi,\tau) \bar w ^{12} (0,0).\Big\rangle 
 \label{(3.22)}\end{equation}

\noindent  Putting all the terms together we have the following 
equation to the lowest order in $\e$:

\begin{align}
\begin{split} \int dx f(x) \rho &[u^\be (x,t) - u^\be (x, 0)] = \\
&\int_0 ^t ds \Big[\int
dy  \sum_{k=1}^d
 {\partial_k f }(y)\{\rho u^\be(y,s) u^k (y,s) -
 \eta
[{\partial_k u^\be }]^{Tr} (y,s)- p-\zeta\sum_{\ga=1}^d{\partial_\ga u^\ga } (y,s)\} \\+&\int dx \sum_{\alpha=1}^d{\partial_{\a\beta\a}^3 f
 }(x)\hat\Phi_{\alpha \beta}(x,s) + \int dx\sum_{k=1}^d
 {\partial_k f }(x)\e^{-1}\Big\langle R_1^s \bar w^{ \be k}\Big\rangle_{G_0}\Big],
 \label{(3.21)}
\end{split}
\end{align} 
for any test function $f$, where $[A]^{Tr}$ means zero trace of $A $.
The differential form is then

$$\partial_t(\rho u)+ \nabla\cdot (u\otimes u)+\nabla p=\nabla\cdot(\tau^{(1)}-\tau^{(2)})$$
$$\tau^{(1)}_{\a\beta}:=\eta\left(\p_\a u^{\beta}+\p_\beta u^{\a}-{2\over d}\d_{\a\beta}\p_\a u^{\a}\right)+\zeta \d_{\a\beta}\p_\a u^{\a}, $$

\begin{equation}\tau^{(2)}_{\a\beta}:=[K_1({\partial_\a }T\partial_\beta T-{1\over d}\sum_{\alpha=1}^d({\partial_\a }T)^2) +{\omega_1 }  \sum_{\alpha=1}^d({\partial_\a }T)^2+ K_2 (\partial^2_{\a\beta}T-{1\over d}\sum_{\alpha=1}^d\partial^2_{\a\alpha}T)+{\omega_2}\sum_{\alpha=1}^d\partial^2_{\a\alpha}T)],\label{tau}
\end{equation}
with
$$K_i=Y_i+Z_i, \quad \omega_i=\bar\omega_i+{\mathcal \phi}_i,\quad i=1,2,$$
computed in the Appendix A.1.

\vskip.2cm

\subsection{\bf Energy equation}
\vskip.2cm

Using the arguments developed 
before  it is possible also to find the equation for the  energy. 

We consider the conservation law for the energy \eqref{(2.9)}.

Therefore we look for the equation for the quantity
$z_i^{d+1}$. By \eqref{(2.9)}we have
\begin{equation}{d\over dt} \e^d\sum_{i=1}^N f(x_i)z_i^{d+1} = \e^{-1}
\e^d \sum_{i=1}^N\sum_{k=1}^d  {\partial_k f }(x_i)
 w^{d+1\, k}_i +O(\e) .\label{(3.24)}\end{equation}
We take the average of both sides w.r.t. $F_\e$. We have 
$$\big\langle   \e^d\sum_{i=1}^N f(x_i)z_i^{{d+1}}\big\rangle_{F_\e}=\int dx f(x)\rho(x) e(x)+ O(\e), $$
and, by using the expression of $F_\e=G_\e+ G_0R_1 +O(\e)$,
$$ \e^{-1}
\e^d \sum_{i=1}^N\sum_{k=1}^d \big\langle  {\partial_k f }(x_i)
 w^{d+1\, k}_i \big\rangle_{F_\e}=
 \int dx {\partial_k f }(x)[\e^{-1}\big\langle  
 w^{d+1\, k}(x)\big\rangle_{G_\e}+\big\langle 
 G_0R_1w^{d+1\, k}(x) \big\rangle]+O(\e).
$$
We need to evaluate
$\big\langle  \e w^{{d+1}\, k}(x)G_0R_1\big\rangle$  and 
$\e^{-1}\big\langle  w^{{d+1}\, k}(x)\big\rangle_{G_\e}$.

The first of them  gives the diffusive
correction. We introduce  $\tilde w ^{{d+1}\, k}$ and 
$\tilde z^{d+1}$ defined 
by  \eqref{(2.4)} with $v$ replaced by $\tilde v$ and%
\begin{equation}w_i ^{{d+1}\, k}=\tilde w_i ^{{d+1}\, k}+ \e \big\{u^k(x_i)
z_i ^{d+1} +\sum_{\ga=1}^d\big( u^\ga  \tilde v_i ^\ga \tilde v_i ^k
+\sum_{j\neq i}\Psi^{\ga k} (\e^{-1}|x_i -x_j|) {1 \over 2}  [u^\ga (x_i) +
u^\ga(x_j)]\big)\big\}. \label{(3.26)}\end{equation}
Then, 
$$\big\langle   \tilde w^{{d+1}\, k}(x){\mathcal{L} ^*} ^{-1}{\mathcal{L}^* } G_0R_1\big\rangle=-\e^{-1}\big\langle   {\mathcal{L}}  ^{-1}\bar w^{{d+1}\, k}(x)\mathcal L^*G_0 \big\rangle,$$
and by \eqref{(2.31.2)} ( only $R_1^a$ enters since  $w^{d+1}$ is odd),
\begin{equation}\Big\langle  \bar w^{{d+1}\, k} G_0R_1^a\Big\rangle=-
\Big\langle \sum_{j=1}^N\sum_{l=1}^d \sum_{\mu =0,4}
{\partial_l \la _0 ^\mu
} (x_j) \bar w^{\mu l}_j {\mathcal{L} } ^{-1}
\bar w^{{d+1}\, k}(x) \Big\rangle_{G_0}+O(\e) ,\label{(3.27)}\end{equation}
where $\bar w^{{d+1}\, k} = \tilde w^{{d+1}\, k} - 
\mathcal{P}(\tilde w^{{d+1}\, k})$.
Because of time-reversal and rotation invariance of the Gibbs state
the only correlations different from zero are (see \cite{S})
\begin{equation}\int dx \Big \langle \bar w ^{{d+1}\, k}(x,\tau)
\bar w^{{d+1}\, l}(0,0)\Big\rangle_{G_0} =\delta_{l k} a(\tau) ,\label{(3.28)}\end{equation} 
and $\int d\tau  a(\tau)=2\kappa T^2$.  Therefore the conductivity
$\kappa$ is given by 
\begin{equation}\kappa={1 \over 2d T^2} \int d\tau \Big\{\int
d\xi \Big \langle\sum_{k,l=1}^d w^{{d+1}\, k}
(x,\tau) w^{{d+1}\, l}(0,0)\Big\rangle -d(T(e
+P)^2/\rho\Big\}. \label{(3.29)}\end{equation} 
We observe that, since $\la_\e=-(T_\e)^{-1}$, $\la_0 ^{d+1}$ is given by
$-T^{-1} $.

Using the previous arguments  one can see that 
the second term in \eqref{(3.26)} gives no contribution to
$\langle w^{{d+1}\,k} R_\e \rangle$ in the limit $\e\to 0$.

The mean of the energy current on the Gibbs
state $G_\e$, i.e. $\langle w^{{d+1}\, k} \rangle_{G_\e}$, 
is nothing but
$(\rho_\e e_\e +P_\e)u_\e$ (see \cite {S}), where 
$ e_\e= \langle z^{d+1}\rangle_{G_\e}$. Hence, $\e^{-1}\langle w^{{d+1}\, k} \rangle_{G_\e}=( \rho e+P)u+o(\e)$
with  $e$  the order zero term for the internal energy and $P$ the order zero term for the pressure.

Summarizing, if we take the average of \eqref{(3.24)} with respect $F_\e$
we obtain 
\begin{align} 
\begin{split} & \int dx f(x)[\rho e(x,t)-\rho e
(x,0)]= \\  & \int_0 ^t ds \int dx f(x) \{-
\text{div}[(\rho e +P)u](x,s)  +\nabla(k \nabla T )(x,s)\}.
 \label{(3.30)}
 \end{split}
 \end{align}
 
 In differential form
 $$\partial_t(\rho e)=-
\text{div}[(\rho e +P)u]+\nabla(k \nabla T).$$

By using the equation for $\rho$ we get also
 $$\rho[\partial_t e+u\cdot\nabla e]=-
P\nabla\cdot u+\nabla(k \nabla T).$$

\subsection{\bf Entropy}
 
 Now we show that these equations  imply the growth in time of the total thermodynamic entropy and hence the second law of the thermodynamics. Let $s(\rho,e)$ denote the entropy as function of the density $\rho$ and the internal energy $e$ and notice that $ {\partial s\over \partial e}={1\over T};\quad {\partial s\over \partial \rho}=-{P\over T\rho^2}$.
We have
 $$\partial_t(\rho s)+\nabla\cdot (\rho u s)=\rho \partial_t s +s\partial_t\rho +\nabla\cdot (\rho u s)=
 {\partial s\over \partial e}\rho\partial_t e+{\partial s\over \partial \rho}\rho\partial_t\rho +s\partial_t\rho+\nabla\cdot (\rho u s)
 $$
 $$
 ={1\over T}\rho\partial_t e-{P\over T\rho}\partial_t\rho- s\nabla\cdot(\rho u)+\nabla\cdot (\rho u s).
 $$
 By using the equation for the energy
 $$\partial_t(\rho s)+\nabla\cdot (\rho u s)=-{1\over T}\rho u\cdot\nabla e-{P\over T}\nabla\cdot u
 +{1\over T}\nabla(k \nabla T)+{P\over T\rho}\nabla\cdot(\rho u)-u\rho\cdot\nabla s
 $$
 $$
 =\nabla ({k\over T}\nabla T)+{1\over T^2}|\nabla T|^2.
 $$
 We have used the identity
 $$
\rho u\cdot[{1\over T}\nabla e+{P\over T\rho^2}\nabla \rho]=\rho u\cdot\nabla s
 $$
 Hence, by integrating over $x$ on a torus
 $$
 \partial_t\int dx \rho s=\int dx {1\over T^2}|\nabla T|^2\ge 0
 $$
 
 \vskip.3cm
\section{Comparison with the Boltzmann case}

It is well known that the incompressible Navier-Stokes-Fourier equations for the perfect gas can be derived from
the Boltzmann equation under a suitable diffusive scaling (scale space as $\e^{-1}$ and time as $\e^{-2}$) and taking  the Mach number  proportional to $\e$ in the limit $\e\to 0$.
Under the same scaling but taking initial conditions with gradient of density and temperature of order $1$ and/or diffusive boundary condition with gradient of temperature of order $1$ the formal limiting equation are different (usually called ghost effect equations following Sone)

\begin{align}\label{fluid system-}
\left\{
\begin{array}{rcl}
    \nabla P&=&\rq\tq,\\\rule{0ex}{1.2em}
    \partial_t{\uq}+\uq\cdot\nabla\uq+\nabla \mathfrak{p}&=&\nabla\cdot\left(\bar\tau^{(1)}-\bar\tau^{(2)}\right),\\\rule{0ex}{1.2em}
    {\partial_ t}\rho+\nabla\cdot(\rq u)&=&0,\\\rule{0ex}{1.8em}
  {3\over 2} \partial_t P(t)+ {5\over 2}P\big(\nabla\cdot u\big) &=&\nabla\cdot\left(\bar\k\dfrac{\nabla\tq}{2\tq^2}\right),
    \end{array}
    \right.
\end{align}
for $d=3$, where  $\bar\k(\tq)>0$ is the heat conductivity,
$$\tau^{(1)}_{ij}:=\lambda \left(\p_iu_{j}+\p_ju_{i}-{2\over3}\d_{ij}\p_iu_{i}\right),$$ $$\tau^{(2)}_{ij}:={\lambda^2\over P}\Big(\bar K_1[\big(\p_i\p_jT\big)-{1\over3}\d_{ij}\sum_i\partial^2_i T]+{\bar K_2\over T}[\big(\p_iT\big)\big(\p_jT\big)-{1\over3}\d_{ij}\sum_i(\partial_iT)^2]\Big )$$ for some smooth function $\lambda[T]>0$, the viscosity coefficient, and positive constants $K_1$ and $K_2$. 

To give the expressions of the transport coefficients define the quantities
\begin{equation}
\ds\m(x, v):={\rho(x)\over\big(2\pi {T}(\vx)\big)^{3/2}}
\exp\bigg(-{\abs{\vv}^2\over 2{T}(\vx)}\bigg)\label{mu},
\end{equation}
\begin{align}\label{final 22}
    &\ab:=v\cdot\left(\abs{v}^2-5T\right)\sqrt\mu\in R^3,\quad {\mathscr{A}}:=\li\left[\ab\right]\in R^3,\\
    &
\end{align}
   $$\bbb=\bigg(v\otimes v-{\abs{v}^2\over 3}\mathbf{1}\bigg)\sqrt\mu\in\r^{3\times3},    \quad \b=L^{-1}\bbb\in\r^{3\times3},$$

\begin{align}\label{final 23}
    \bar kI:=\int_{R^3}\left( {\mathscr{A}}\otimes\ab\right)\ud v,\quad \lambda:={1\over \tq}\int_{R^3}\b_{ij}\bbb_{ij}\ \ \text{for}\ \ i\neq j.
\end{align}
where $Lf{1\over\sqrt\mu}=\lc (\sqrt\mu f)$ and $\lc $ is the linearized Boltzmann operator around the local Maxwellian $\mu$ with zero mean for hard sphere,
$$
{\lambda^2\over P}\bar K_1={1\over T^2}\int dv\mathscr{B}_{ij}v_i {\mathscr{A}}_j,\quad i\ne j, 
$$

$$
{\lambda^2\over  TP}\bar K_2={1\over T^4}\int dv\mathscr{B}_{ij}\Big[\Gamma({\mathscr{A}}_i,{\mathscr{A}}_j)+v_i\sqrt\mu^{-1}{\partial\over \partial T}({\sqrt\mu\over T^2}{\mathscr{A}}_j)\Big]
\quad i\ne j$$
where $\Gamma$ is the collision operator in the Boltzmann equation.

We want to compare the equations we get for the Hamiltonian particle system and the ones derived from Boltzmann, taking into account that the state equation in the second case is the one for the perfect gas $P(\rho,T)=\rho T$, since
the Boltzmann equation describes a rarefied gas.
The continuity equation in kinetic theory is the same as for particle system and also the condition 
$\nabla P=0$.

The equation for the energy if we consider the state equation of a perfect case becomes:

For a perfect gas $e={3\over 2}T$ and 
 $$
{3\over 2} \partial_t P(t)=-
{5\over 2}P\nabla\cdot u+\nabla(k \nabla T).
 $$
 If the domain is a torus, $\partial_tP(t)=0$ and we get
 $${5\over 2}P\nabla\cdot u=\nabla(k \nabla T).$$
 
 If   we compare the equation for the momentum we see that  also in this case the structure of the equation is the same included the new thermal stress terms.
 
 Obviously, the transport coefficients are different and we are not able to compare them, since the Boltzmann equation is modelling a gas of hard sphere and for the particles we were considering a smooth potential. However, we notice that
 if we represent  ${\mathcal L}^{-1}$ as
$${\mathcal L}^{-1}f=\int_0^\infty ds e^{-s{\mathcal L}} f(s)$$
the transport coefficients $\lambda$ and $\bar \k$ are also in kinetic theory expressed as time correlations of currents \begin{equation}
\lambda=-{1\over T}\int_0^\infty ds <(|v_i v_j -{1\over 3}|v|^2 I)(s)(|v_i v_j -{1\over 3}|v|^2
I)(0)>_{\mu},
\end{equation}
\begin{equation}
\bar \k=-\int_0^\infty ds<{1\over
2}(|v|^2-5 T)v_i(s) {1\over
2}(|v|^2-5 T)v_i(0)>_{\mu}.
\end{equation}

 Notice that $\zeta $ in the kinetic theory is zero and we realise that also $\omega_1$ and $\omega_2$ are not present in kinetic theory. The expression of the transport coefficients as given by the Green-Kubo formulas are similar with the difference that in kinetic theory is present the linearized Boltzmann operator 
 around the local Maxwellian equilibrium while for the particles the same role is played by the Liouville operator. Moreover, the averages are taken versus the local Maxwellian equilibrium with zero velocity $\mu$
 in kinetic theory and versus the {\it local} Gibbs equilibrium $G_0$ for particles. 
 
 Moreover, in kinetic theory $\lambda$ and $\bar \k$ are well defined due to the hypocoericvity property of $L$, while for the particle system we need ergodic and mixing properties of  the Liouville operator 
${\mathcal L}$.

 On the other hand, the new transport coefficients $K_1$ and $K_2$ have two different contributions: one involving spatial correlations, which are not present in kinetic theory, and one involving space and double time correlations which are similar in this respect to the ones present in kinetic theory. 
 
 To conclude , 
we remark that also in the Boltzmann case the entropy grows in time as showed by Bobylev \cite{Bo}.

 Finally, we want to stress that the argument for the particles is completely formal and some rigourous results in the case of the incompressible Navier-Stokes equations can be obtained for stochastic systems of particles on the lattice \cite{EMY},\cite{EMY1}.  The model of stochastic termal gas on the lattice in \cite{BEM} could be useful for deriving the new equations.
 
 In the case of the kinetic theory the ghost effect equations were obtained in one-dimensional stationary cases, \cite{AEMN2}, \cite{Brull2008}, \cite{ELM}. A recent advance has been obtained  in the stationary case in a general domain for a rarefied gas in contact with a thermal reservoir  with non-homogeneous temperature: the rigourous proof of the hydrodynamic limit \cite {EGMW0} \cite{EGMW}.

\newpage

\appendix
\makeatletter
\renewcommand \theequation {%
A.
\ifnum\c@subsection>\z@\@arabic\c@subsection.%
\fi
\@arabic\c@equation} \@addtoreset{equation}{section}
\@addtoreset{equation}{subsection} \makeatother
 \section{}

 \subsection{ Transport coefficients}
\vskip.01 cm

 We examine more closely the expression of the transport coefficients $K_1, K_2$ in \eqref{tau} and first we compute $Y_1,Y_2$ in \eqref{Y} and then the part $Z_i, i=1,2$ due to the term \eqref{RS}.
 
  $\bullet$ {\it Computation of $Y_i,i=1,2$. }
  
 $Y_2(x,t)={\partial\over \partial T}\hat\Phi_{ \alpha\beta}(x,t), \alpha\ne \beta$ is defined in   \eqref{YY}. 
 We have
 $$\hat\Phi_{ \alpha\beta}:=\Big\langle\bar \Phi_{ \alpha\beta}\Big\rangle _{G_0}=
\Big\langle{G_0}(x,t)\bar\Phi_{ \alpha\beta}(x)\Big\rangle 
$$
where 
$$\bar\Phi_{ \alpha\beta}(x)={1 \over 2}
\e^d\sum_{i,j=1,i\ne j}^N\delta(x-x_i)\partial_\be V (\e^{-1}(x_i -x_j))[\xi_i ^\beta  -\xi_j^\beta ][\xi_i ^\a 
-\xi_j^\a ]^2. $$ We put ${\partial\over \partial T}\lambda_0^\mu=(\lambda_0^\mu)'$ and  ${\partial^2\over \partial T^2}\lambda_0^\mu=(\lambda_0^\mu)''$ for  $\mu=0,d+1$. Then,

$$Y_2=\Big\langle{\partial G_0\over \partial T}{}\bar\phi_{\alpha \beta}\Big\rangle =
\Big\langle\sum_{k=1}^Nz^0_k(\lambda_0^0)'(x_k)\bar\Phi_{\alpha \beta}\Big\rangle_{G_0}-\Big\langle\sum_{k=1}^Nz^0_k(\lambda_0^0)'(x_k)\Big\rangle_{G_0}\Big\langle\bar\Phi_{\alpha \beta}\Big\rangle_{G_0}
$$
$$+ 
\Big\langle\sum_{k=1}^Nz^{d+1}_k(\lambda_0^{d+1})'(x_k)\bar\Phi_{\alpha \beta}\Big\rangle_{G_0}-\Big\langle\sum_{k=1}^Nz^{d+1}_k(\lambda_0^{d+1})'(x_k)\Big\rangle_{G_0}\Big\langle\bar\Phi_{\alpha \beta}\Big\rangle_{G_0}.
$$
Now call $ (\tilde z^\mu)=z^\mu-\Big\langle z^\mu\Big\rangle_{G_0}$ and $\tilde \Phi_{\alpha \beta}=\bar \Phi_{\alpha \beta}-\Big\langle\bar\Phi_{\alpha \beta}\Big\rangle_{G_0}$. Then,
$$Y_2=\Big\langle\sum_{k=1}^N\tilde z^0_k(\lambda_0^0)'(x_k)\tilde\Phi_{\alpha \beta}\Big\rangle_{G_0}+\Big\langle\sum_{k=1}^N\tilde z^{d+1}_k(\lambda_0^{d+1})'(x_k)\tilde\Phi_{\alpha \beta}\Big\rangle_{G_0}.
$$
Let $h(x)$ be a test function and compute $\int dx h(x)Y_2(x)$.
$$\int dx h(x)Y_2(x)=\e^{-d}\sum_{\mu=0,4}\int dx h(x)\int dy(\lambda_0^\mu)'(y)\Big\langle\tilde z^\mu(y)\tilde\Phi_{\alpha \beta}(x)\Big\rangle_{G_0}.$$
We change variable $x=y+\e\zeta$ to absorb $\e^{-d}$ and get
$$
\int dx h(x)Y_2(x)=\sum_{\mu=0,4}\int dy h(y)(\lambda_0^\mu)'(y)\int d\zeta\Big\langle\tilde z^\mu(0)\tilde\Phi_{\alpha \beta}(\zeta)\Big\rangle_{G_0}.
$$
In conclusion
$$
Y_2(y)=\sum_{\mu=0,4}(\lambda_0^\mu)'(y)\int d\zeta\Big\langle\tilde z^\mu(0)\tilde\Phi_{\alpha \beta}(\zeta)\Big\rangle_{G_0}.
$$

For $\alpha\ne \beta$
$$Y_1={\partial^2 \hat\Phi_{\alpha\beta}\over \partial T^2}={\partial Y_2\over \partial T}=\sum_{\mu=0,4}(\lambda_0^\mu)''(y)\int d\zeta\Big\langle\tilde z^\mu(0)\tilde\Phi_{\alpha \beta}(\zeta)\Big\rangle_{G_0}$$
$$
+\sum_{\mu, \nu=0,4}\int dy(\lambda_0^\mu)'(y){(\lambda_0^\nu)}'(y)\int d\zeta\int d\zeta'\Big\langle\tilde z^\mu(0)\tilde z^{\nu}(\zeta')\tilde\Phi_{\alpha \beta}(\zeta)\Big\rangle_{G_0}.
$$

  $\bullet$ {\it Computation of $Z_i,i=1,2$. }
  
\noindent Now we compute the contribution to the transport coefficients due to the term
$$C:=\e^{-1}\Big\langle R_1^s\e^d \sum_{i=1}^N \sum_{k=1}^d\bar w^{ \be k}_i{\partial_x f }(x_i)\Big\rangle_{G_0}.$$
We use the expression of $R_1^s$ given by \eqref{(2.31.1)}
$$C=\e^{-1}\Big\langle G_0R_1^s\e^d \sum_{i=1}^N \sum_{k=1}^d\bar w^{ \be k}_i{\partial_k f }(x_i)\Big\rangle$$
$$=\e^{-2}\Big\langle{\mathcal{L}^* } ^{-1}{\mathcal{L}^* } ^{-1}[{\mathcal{L}^* } {\mathcal{L} ^*}G_0] \e^d \sum_{i=1}^N \sum_{k=1}^d\bar w^{ \be k}_i]{\partial_k f }(x_i)\Big\rangle$$
$$
=\e^{-1}\Big\langle{\mathcal{L} ^*} [G_0\sum_{\ell=1}^N \sum_{\mu=0,4}\sum_{\ga=1}^d{\partial_\gamma \lambda_0^\mu }(x_{\ell})w^{\mu\gamma}_{\ell}] {\mathcal{L} } ^{-1}{\mathcal{L} } ^{-1}[\e^d \sum_{i=1}^N\sum_{k=1}^d\bar w^{ \be k}_i{\partial_k f }(x_i)]\Big\rangle.
$$

We compute now 
$$A:=\e^{-1}[{\mathcal{L} ^*} G_0]\sum_{\ell=1}^N \sum_{\mu=0,4}\sum_{\ga=1}^d{\partial_\gamma \lambda_0^\mu  }(x_{\ell})w^{\mu\gamma}_{\ell}.$$
\begin{align}
\begin{split}
&A=[\e^{-1}{\mathcal{L} ^*} G_0]\sum_{\ell=1}^N \sum_{\mu=0,4}\sum_{\ga=1}^d{\partial_\gamma \lambda_0^\mu }(x_{\ell})w^{\mu\gamma}(x_{\ell})+\e^{-1}G_0{\mathcal{L} ^*}[\sum_{\ell=1}^N \sum_{\mu=0,4}\sum_{\ga=1}^d{\partial_\gamma \lambda_0^\mu }(x_{\ell})w^{\mu\gamma}(x_{\ell})]
\\
&= G_0[\sum_{\ell=1}^N \sum_{\mu=0,4}\sum_{\ga=1}^d{\partial_\gamma \lambda_0^\mu }_{\ell}w^{\mu\gamma}_{\ell}][\sum_{s=1}^N \sum_{\mu=0,4}\sum_{\nu=1}^d{\partial_\nu \lambda_0^\mu }(x_{s})w^{\mu\nu}_{s}]\\
&+\e^{-1}G_0\sum_{\ell=1}^N \sum_{\mu=0,4}\sum_{\ga=1}^d{\mathcal{L} ^*}[{\partial_\gamma \lambda_0^\mu }(x_{\ell})w^{\mu\nu}_{\ell}].
\label{(A.1)}
\end{split}
\end{align}
The first term in the rhs of \eqref{(A.1)} is
$$
\e\Big[\sum_{\mu=0,4}\sum_{\ga=1}^d\e^{-d}\int dx (\lambda_0^\mu)'\sum_{\ga=1}^d\partial_\gamma T(x)w^{\mu\gamma}(x)\quad \e^{-d}\int dy (\lambda_0^\mu)'\sum_{\nu=1}^d\partial_\nu T(y)w^{\mu\nu}(y)
$$
$$
+2\e^{-d}\int dx (\lambda_0^0)'\sum_{\ga=1}^d\partial_\gamma T(x)w^{0\gamma}(x)\quad \e^{-d}\int dy (\lambda_0^{d+1})'\sum_{\nu=1}^d\partial_\nu T(y)w^{d+1\nu}(y)\Big].
$$
The contribution to $C$ is
$$
\sum_{\mu=0,4}\sum_{\ga,\nu=1}^d\e^{-d}\int dx (\lambda_0^\mu)'\partial_\gamma T(x)\e^{-d}\int dy (\lambda_0^\mu)'\partial_\nu T(y)
$$
$$\times\int dz\partial_kf(z)
\int_0^\infty ds\int_0^\infty dt\Big\langle w^{\mu\gamma}(x 0)w^{\mu\nu}(y 0)\bar w^{\beta k}(z s+t)\Big\rangle_{G_0}
$$
$$
+2\e^{-d}\sum_{\ga,\nu=1}^d\int dx (\lambda_0^0)'\partial_\gamma T(x)\e^{-d}\int dy (\lambda_0^{d+1})'\partial_\nu T(y)$$
$$\times\sum_{k=1}^d\int dz\partial_kf(z)
\int_0^\infty ds\int_0^\infty dt\Big\langle w^{0\gamma}(x 0)w^{d+1\nu}(y 0)\bar w^{\beta k}(z s+t)\Big\rangle_{G_0}.
$$
By the change of  variables $x=z+\e\xi,\quad y=z+\e\xi',\quad s+t=\tau$
$$
=\sum_{\mu=0,d+1}\sum_{\ga,\nu,k=1}^d\int dz\partial_kf(z)(\lambda_0^\mu)'\partial_\gamma T(z)(\lambda_0^\mu)'\partial_\nu T(z)
$$
$$\times\int_0^\infty ds\int_s^\infty d\tau\int d\xi \int d\xi' \Big\langle w^{\mu\gamma}(\xi \tau)w^{\mu\nu}(\xi' \tau)\bar w^{\beta k}(0 0)\Big\rangle_{G_0}
$$
$$
+2\sum_{\ga,\nu,k=1}^d\int dz\partial_kf(z)(\lambda_0^0)'\partial_\gamma T(z)(\lambda_0^{d+1})'\partial_\nu T(z)
$$
$$\times\int_0^\infty ds\int_s^\infty d\tau\int d\xi \int d\xi' \Big\langle w^{0\gamma}(\xi \tau)w^{d+1\nu}(\xi' \tau)\bar w^{\beta k}(0 0)\Big\rangle_{G_0}.
$$
These terms give rise in the equation for $u^\beta$ to terms of the form 
$$
-2\sum_{\ga,\nu,k=1}^d\partial_k\Big[[(\lambda_0^0)'\partial_\gamma T(z)(\lambda_0^{d+1})'\partial_\nu T(z)] \a^{0 d+1}_{\gamma\nu\beta k}\Big ]
$$
and for $\mu=0,d+1$ 
$$
-\sum_{\ga,\nu,k=1}^d\partial_k\Big[[(\lambda_0^\mu)'\partial_\gamma T(z)(\lambda_0^{\mu})'\partial_\nu T(z)] \a^{\mu \mu}_{\gamma\nu\beta k}\Big ],
$$
where 
$$\a^{0 d+1}_{\gamma\nu\beta k}:=\int_0^\infty ds\int_s^\infty d\tau\int d\xi \int d\xi' \Big\langle w^{0\gamma}(\xi \tau)w^{d+1\nu}(\xi' \tau)\bar w^{\beta k}(0 0)\Big\rangle_{G_0}
$$

$$=a_1(t,z)\delta_{\beta k}\delta_{\gamma \nu}+a_2(t z)\delta_{\beta \nu}\delta_{\gamma k}+a_3(t,z)\delta_{\beta \gamma}\delta_{\nu k},
$$

$$\a^{0 0}_{\gamma\nu\beta k}
:=\int_0^\infty ds\int_s^\infty d\tau\int d\xi \int d\xi' \Big\langle w^{0\gamma}(\xi \tau)w^{0\nu}(\xi' \tau)\bar w^{\beta k}(0 0)\Big\rangle_{G_0}
$$ $$
=b_1(t,z)\delta_{\beta k}\delta_{\gamma \nu}+b_2(t z)\delta_{\beta \nu}\delta_{\gamma k}+b_3(t,z)\delta_{\beta \gamma}\delta_{\nu k},
$$
$$\a^{d+1 d+1}_{\gamma\nu\beta k}
:=\int_0^\infty ds\int_s^\infty d\tau\int d\xi \int d\xi' \Big\langle w^{d+1\gamma}(\xi \tau)w^{d+1\nu}(\xi' \tau)\bar w^{\beta k}(0 0)\Big\rangle_{G_0}
$$ $$
=c_1(t,z)\delta_{\beta k}\delta_{\gamma \nu}+c_2(t z)\delta_{\beta \nu}\delta_{\gamma k}+c_3(t,z)\delta_{\beta \gamma}\delta_{\nu k}
$$
Since the potential is central we have $a_2=a_3$, $b_2=b_3$ $c_2=c_3$ \cite{S}. 

Now we compute the second term in \eqref{(A.1)} $$B:=\e^{-1}G_0\sum_{\ell=1}^d \sum_{\mu=0,d+1}\sum_{\ga=1}^d{\mathcal{L} ^*}[{\partial_\gamma \lambda_0^\mu }(x_{\ell})w^{\mu\gamma}_{\ell}].
$$
We have for

$\mu=0:$
$$\quad\quad\e^{-1}\sum_{\ell=1}^d \sum_{\mu=0,4}\sum_{\ga=1}^d {\mathcal{L} ^*}[{\partial_\gamma \lambda_0^0 }(x_{\ell})w^{0\gamma}_{\ell}]=\e^{-d}\sum_{\gamma,\nu=1}^d\int dx \partial^2_{\nu\gamma}\lambda_0^0(x)w^{\nu\gamma}(x).
$$
The corresponding term in the equation for $u^\beta$ is 
$$-\sum_{\ga,\nu,k=1}^d \nabla_k\Big[[(\lambda_0^0)'\partial^2_{\nu\gamma} T(z)+(\lambda_0^0)''\partial_\gamma T(z)\partial_\nu T(z)] h^{0 }_{\gamma\nu\beta k}\Big ],
$$
with
$$h^{0 }_{\gamma\nu\beta k}=\int_0^\infty ds\int_s^\infty d\tau\int d\xi  \Big\langle \bar w^{\nu\gamma}(\xi \tau)\bar w^{\beta k}(0 0)\Big\rangle_{G_0}=h_1(t,z)\delta_{\beta k}\delta_{\gamma \nu}+h_2(t z)\delta_{\beta \nu}\delta_{\gamma k}+h_3(t,z)\delta_{\beta \gamma}\delta_{\nu k}.
$$
The most difficult term is

 $\mu=d+1$:
$$
\quad \sum_{\ell=1}^N \sum_{\ga=1}^d {\mathcal{L} ^*}[{\partial_\gamma \lambda_0^{d+1} ( x_{\ell})}w^{d+1\gamma}_{\ell}]:=\sum_{\ell=1}^N  \sum_{\ga=1}^d{\mathcal{L} ^*}[ g_\gamma( x_{\ell})w^{d+1\gamma}_{\ell}],$$
where, for semplicity, we put $g_\gamma(x)=\partial_\gamma \lambda_0^{d+1} ( x)$.
$$
{\mathcal H}:=\sum_{\ell=1}^N  \sum_{\ga=1}^d{\mathcal{L} ^*}[ g_\gamma( x_{\ell})w^{d+1\gamma}_{\ell}]$$
\begin{equation}
=-\sum_{\ell=1}^N  \sum_{\ga=1}^d\Big \{\e\sum_{k=1}^dv_\ell^k\partial_{ k}g_\gamma( x_{\ell})w^{d+1\gamma}_{\ell}
+\sum_{i=1}^N \sum_{k=1}^d v_i^k g_\gamma( x_{\ell}){\partial\over\partial x_i^k}w^{d+1\gamma}_{\ell}\Big\}
\label{1}
\end{equation}
\begin{equation}
-\sum_{\ell=1}^N  \sum_{\ga=1}^dg_\gamma( x_{\ell})\Big \{\sum_{i=1}^N \sum_{k=1}^d\sum_{j\ne i} \partial_k V(\e^{-1}(x_i-x_j)){\partial \over \partial v_i^k} w^{d+1\gamma}_{\ell}
\Big\}.\label{2}
\end{equation}
The first term is of order $\e$ and will appear in $C$. The others has to be examined and we start with the second term in \eqref{1}
$$-\sum_{\ell=1}^N  \sum_{\ga=1}^d\sum_{i=1}^N \sum_{k=1}^d v_i^k g_\gamma( x_{\ell}){\partial\over\partial x_i^k}w^{d+1\gamma}_{\ell}
$$
\begin{equation}=-\sum_{\ell=1}^N  \sum_{\ga=1}^d
\sum_{k=1}^d g_\gamma( x_{\ell})\Big[{1\over 2}\sum_j\sum_\nu(\partial_k\Psi^{\nu\ga}(\e^{-1}(x_\ell-x_j))(v_j^k-v_\ell^k){1\over 2}(v_\ell^\nu+v_j^\nu)
+\sum_i{1\over 2}v_\ell^\ga v_i^k{\partial\over\partial x_i^k}z_\ell^{d+1\ga}\Big].
\label{9}
\end{equation}
The first  term will not give contribution to $C$ because of the Gaussian integration on the velocities. Now we pass to the  term in \eqref{2} and add the last term in in \eqref{9}
\begin{align}
\begin{split}&\sum_{\ell=1}^N  \sum_{\ga=1}^dg_\gamma( x_{\ell})\sum_{i=1}^N \Big \{\sum_{k=1}^d\sum_{j\ne i} \partial_k V(\e^{-1}(x_i-x_j)){\partial \over \partial v_i^k}w^{d+1\ga}_\ell+{1\over 2}v_i^kv_\ell^\ga{\partial\over\partial x_i^k}z_\ell^{d+1\ga}
\Big\}\\
=&\sum_{\ell=1}^N  \sum_{\ga=1}^dg_\gamma( x_{\ell})\sum_{i=1}^N\Big \{ {1\over 2}\sum_{k=1}^dv_i^kv_\ell^\ga{\partial\over\partial x_i^k}z_\ell^{d+1\ga}
+\sum_{k=1}^d\sum_{j\ne i} \partial_k V(\e^{-1}(x_i-x_j))\Big[z^{d+1}_\ell{\partial \over \partial v_i^k} v_\ell ^\ga\\
&+v_\ell ^\ga{\partial \over \partial v_i^k}{1\over 2}|v_\ell |^2+{1\over 2}\sum_{s=1}^N  \sum_{\nu=1}^d\Psi^{\nu\gamma}(\e^{-1}(x_\ell-x_s)
{1\over 2}{\partial \over \partial v_i^k}[v_\ell^\nu+v_s^\nu])\big]\Big\}.
\label{(1.2)} 
\end{split}
\end{align}
We begin to deal with the second term  in the rhs of  \eqref{(1.2)} 
$${\mathcal M}:=\sum_{\ell=1}^N  \sum_{\ga=1}^dg_\gamma( x_{\ell})z^{d+1}_\ell \sum_{j\ne \ell} \partial_\ga V(\e^{-1}(x_\ell-x_j)) .$$

By using the antisymmetry of the gradient of the potential under the exchange $\ell\to j$ we get
$$
{\mathcal M}={1\over 2}\sum_{\ell=1}^N  \sum_{\ga=1}^d\sum_{j\ne\ell,1}^N[g_\gamma( x_{\ell})z_\ell^{d+1}-g_\gamma( x_{j})z_j^{d+1}]{\partial_\ga V}(\e^{-1}(x_\ell-x_j))
$$
$$={1\over 2}\sum_{\ell=1}^N  \sum_{\ga=1}^d\sum_{j\ne\ell,1}^n\Big[[g_\gamma( x_{\ell})-g_\gamma( x_{j})]z_\ell^{d+1}+g_\gamma( x_{j})[z_\ell^{d+1}-z_j^{d+1}]{\partial_\ga V}(\e^{-1}(x_\ell-x_j))\Big]
$$
$$
={1\over 2}\sum_{\ell=1}^N  \sum_{\ga=1}^d\sum_{j\ne\ell,1}^N\Big[[g_\gamma( x_{\ell})-g_\gamma( x_{j})]z_\ell^{d+1}+{1\over 2}[g_\gamma( x_{j})+g_\gamma( x_{\ell})[z_\ell^{d+1}-z_j^{d+1}]{\partial_\ga V}(\e^{-1}(x_\ell-x_j))\Big].
$$
As usual, we replace $[g_\gamma( x_{\ell})-g_\gamma( x_{j})]$ with $\e\partial_sg_\gamma( x_{\ell})(\xi^s_\ell-\xi^s_j)$  so that the first term is of order $\e$ and will give contribution to the equation, while the second term is not of order $\e$ and we have to show that does not give contribution.
$$
{\mathcal M}={1\over 2}\sum_{\ell=1}^N  \sum_{\ga=1}^d\sum_{j\ne\ell,1}^N\Big[\e\sum_{s=1}^d[z_\ell^{d+1}\partial_s g_\gamma( x_{\ell})\Psi^{\gamma s}(\e^{-1}(x_\ell-x_j))
$$
\begin{equation}
+{1\over 4}[g_\gamma( x_{j})+g_\gamma( x_{\ell})][z_\ell^{d+1}-z_j^{d+1}]\partial _\gamma V(\e^{-1}(x_\ell-x_j))\Big]\label{giusto?}
\end{equation}
 In $z_k^{d+1}={1\over 2}\big[ v^2 _k +\sum_{j \ne i=1}^N
 V(\e^{-1}|x_k -x_j|)$ the second term $\sum_{j \ne i=1}^NV(\e^{-1}|x_k -x_j|)$ is 
 \begin{equation}
  \sum_{\ell, j=1}^N-V(\e^{-1}(x_j-x_t))]\partial_\ga V(\e^{-1}(x_\ell-x_j))[g_\ga(x_j)+g_\ga(x_\ell)]\label{zero?}
\end{equation}
 and will not give contribution to $C$ since the term $\sum_{t=1}^N [V(\e^{-1}(x_\ell-x_t))$ under the average on the position in $C$ will not depend on $\ell$ by isotropy.

 Then, the second line in \eqref{giusto?} becomes 
 $${1\over 2}\sum_{\ell=1}^N  \sum_{\ga=1}^d\sum_{j\ne\ell,1}^n{1\over 4}[g_\gamma( x_{j})+g_\gamma( x_{\ell})]{1\over 2}[|v_\ell |^2-|v_j|^2]{\partial_\ga V}(\e^{-1}(x_\ell-x_j)).$$

This term will not give contribution to $C$ because of the Gaussian integration on the velocities. Now, we  compute the first and third   terms  term in \eqref{(1.2)}.

\begin{equation}
\sum_{\ell=1}^N  \sum_{\ga=1}^dg_\gamma( x_{\ell})\sum_{i=1}^N\Big \{ \sum_{k=1}^d\sum_{j\ne i} \partial_k V(\e^{-1}(x_i-x_j))v_\ell ^\ga{\partial \over \partial v_i^k}{1\over 2}|v_\ell |^2+{1\over 2}v_i^kv_\ell^\ga{\partial\over\partial x_i^k}z_\ell^{d+1\ga}\end{equation}

$$
=\sum_{\ell=1}^N \sum_{k=1}^dg_\ga(x_{\ell}){1\over 2}\biggl\{-2v_\ell^\gamma v_\ell^{k}\sum_{j:j\ne \ell}^{1,N}{\partial _k V}(\e^{-1}(x_\ell-x_j))+\sum_{j:j\ne \ell}^{1,N}v_\ell^\gamma\partial_k V
(\e^{-1}(x_\ell-x_j))(v_\ell^{k}-v_j^{k})\biggr\}
$$
$$=-\sum_{\ell=1}^N \sum_{k=1}^dg_\ga(x_{\ell}){1\over 2}\sum_{j:j\ne \ell}^{1,N}v_\ell^\gamma \partial_k V
(\e^{-1}(x_\ell-x_j))(v_\ell^{k}+v_j^{k}).
$$

By using the antisymmetry of $\partial_k V$ we get
$$-{1\over 4}\sum_{\ga=1}^d\sum_{\ell=1}^N[g_\ga(x_{\ell})v_\ell^\gamma-g_\ga(x_{j})v_j^\gamma]\{\sum_{j:j\ne \ell}^{1,N}\sum_{k=1}^d\partial_k V
(\e^{-1}(x_\ell-x_j))(v_\ell^{k}+v_j^{k})\}
$$
$$
=-{1\over 4}\sum_{\ell=1}^N\sum_{\ga=1}^d\Big[[v_\ell^\gamma[g_\ga(x_{\ell})-g_\ga(x_{j})]+g_\ga(x_{j})[v_\ell^\gamma-v_j^\gamma]\Big]\{\sum_{j:j\ne \ell}^{1,N}\sum_{k=1}^d\partial_k V
(\e^{-1}(x_\ell-x_j)){1\over 2}(v_\ell^{k}+v_j^{k})\}
$$
\begin{equation}
=-{1\over 4}\sum_{\ell=1}^N\sum_{\ga=1}^d\Big[\e v_\ell^\gamma\sum_{\nu=1}^d\partial_\nu g_\ga(x_{\ell})\sum_{j:j\ne \ell}^{1,N}\sum_{k=1}^d\Psi^{\nu k}
(\e^{-1}(x_\ell-x_j)){1\over 2}(v_\ell^{k}+v_j^{k})\Big]\label{10}
\end{equation}
$$+g_\ga(x_{j})[v_\ell^\gamma-v_j^\gamma]\Big]\{\sum_{j:j\ne \ell}^{1,N}\sum_{k=1}^d\partial_k V
(\e^{-1}(x_\ell-x_j)){1\over 2}(v_\ell^{k}+v_j^{k})\}.
$$
The first term is of order $\e$ and will appear in $C$ while the second term is zero after average on the velocities in $C$.

Finally, we discuss the last term in \eqref{(1.2)}.
$$\sum_{\ell=1}^N  \sum_{\ga=1}^dg_\gamma( x_{\ell})\sum_{k=1}^d\sum_{j\ne i} \partial_k V(\e^{-1}(x_i-x_j)){1\over 2}\sum_{s=1}^N  \sum_{\nu=1}^d\Psi^{\nu\gamma}(\e^{-1}(x_\ell-x_s)
{1\over 2}{\partial \over \partial v_i^k}[v_\ell^\nu+v_s^\nu])\big]$$
$$=\sum_{\ell=1}^N  \sum_{\ga=1}^dg_\gamma( x_{\ell}){1\over 4}\sum_{s=1}^N \sum_{k=1}^d\sum_{j} [\partial_k V(\e^{-1}(x_\ell-x_j))+\partial_k V(\e^{-1}(x_s-x_j))]\Psi^{k\gamma}(\e^{-1}(x_\ell-x_s)
\big]
$$
The first term is equal to
$$
={1\over 8}\sum_{\ell,j=1}^N  \sum_{\ga,k=1}^d\sum_{s=1}^N \partial_k V(\e^{-1}(x_\ell-x_j))\{[g_\gamma( x_{\ell})\Psi^{k\gamma}(\e^{-1}(x_\ell-x_s)-g_\gamma( x_{j})\Psi^{k\gamma}(\e^{-1}(x_j-x_s)]$$
$$={1\over 8}\sum_{\ell,j=1}^N  \sum_{\ga,k=1}^d\sum_{s=1}^N  [\partial_k V(\e^{-1}(x_\ell-x_j))[g_\gamma( x_{\ell})-g_\gamma( x_{j})]\Psi^{k\gamma}(\e^{-1}(x_\ell-x_s)
$$
$$+g_\gamma( x_{j}[\Psi^{k\gamma}(\e^{-1}(x_\ell-x_s))-\Psi^{k\gamma}(\e^{-1}(x_j-x_s))]\}
$$
The first term is of order $\e$ and the second
  has the same symmetry properties as the term dealt with in \eqref{zero?} and hence does not give contribution.
The other term is 
$$\sum_{\ell,j=1}^N  \sum_{\ga,k=1}^dg_\gamma( x_{\ell}){1\over 4}\sum_{s=1}^N  \partial_k V(\e^{-1}(x_s-x_j))\Psi^{k\gamma}(\e^{-1}(x_\ell-x_s))
$$
$$
=\sum_{\ell,j=1}^N  \sum_{\ga,k=1}^dg_\gamma( x_{\ell})\sum_{s=1}^N  \partial_k V(\e^{-1}(x_s-x_j))[\Psi^{k\gamma}(\e^{-1}(x_\ell-x_s))-\Psi^{k\gamma}(\e^{-1}(x_\ell-x_j))]
$$
and does not give contribution to $C$ for the same reasons.

Putting all the terms of order $\e$  together in \eqref{1},\eqref{giusto?} and \eqref{10} we have that  
$$
\e\e^{-d}\int dx \sum_{s,\ga=1}^d\partial_s g_\gamma(x)(v^\gamma w^{d+1s})(x)+\e\e^{-d}\int dx\sum_{m,\ga=1}^d\partial_mg_\gamma(x){\mathcal A}^{\gamma m}(x)
$$
$$+\e\e^{-d}\sum_{s,\ga=1}^d\int dx \partial_s g_\ga( x){\mathcal C}^{\ga s}(x)+\e\e^{-d}\sum_{s,\ga=1}^d\int dx \partial_s g_\ga( x){\mathcal N}^{\ga s}(x),
$$
with
$${\mathcal C}^{\ga s}(x)=\sum_{\ell, j=1}^N\delta(x_\ell-x){1\over 4}\sum_{\nu,\ga=1}^d\Psi^{\nu\gamma}(\e^{-1}(x_\ell-x_j))v_\ell^s{1\over 2}[v_\ell^\nu+v_j^\nu],
$$
$$ {\mathcal A}^{\gamma m}(x)=\sum_\ell\delta(x-x_\ell)z_\ell^{d+1}\sum_s\Psi^{\gamma m}(\e^{-1}(|x_\ell-x_s|),$$
$$ {\mathcal N}^{\gamma s}(x)=\sum_\ell\delta(x-x_\ell)\sum_{s,j}\partial_m V(\e^{-1}(x_\ell-x_j))\Psi^{s\gamma m}(\e^{-1}(|x_\ell-x_s|).$$

\vskip.1cm 
The contribution to $C$ is then
$$\sum_{m,k,\ga=1}^d\int dz \partial_kf(z)\Big[\partial^2_{m\gamma}(\lambda_0^{d+1})(z)\int_0^\infty ds\int_s^\infty d\tau\int d\xi   \Big\langle(v^\gamma w^{d+1m}+{\mathcal A}^{\gamma m}(x))(\xi,\tau)\bar w^{\beta k}(0,0)\Big\rangle_{G_0}\Big ]
$$
$$
=\sum_{m,k,\ga=1}^d\int dz \partial_kf(z)\sum_m\Big[(\lambda_0^{d+1})''\partial_\gamma T\partial_m T+(\lambda_0^{d+1})'\partial^2_{m\gamma}T(z) d_{\gamma\mu\beta k}\Big ],
$$
with
$$
d_{\gamma\mu\beta k}:=\int_0^\infty ds\int_s^\infty d\tau\int d\xi   \Big\langle([v^\gamma w^{d+1m}+{\mathcal A}^{\gamma m}(x) ](\xi,\tau)]\bar w^{\beta k}(0,0)\Big\rangle_{G_0}
$$
$$
=d_1(t,z)\delta_{\beta k}\delta_{\gamma \mu}+d_2(t z)\delta_{\beta \mu}\delta_{\gamma k}+d_3(t,z)\delta_{\beta \gamma}\delta_{\mu k};\quad d_2=d_3
$$
and for the term ${\mathcal C}$: 
 $\e^{-d}\e\int dx  {\mathcal C}^{\ s\gamma}(x)
\partial_s g_\ga( x) 
$ in the equation

$$
\partial_k[((\lambda_0^{d+1})'\partial^2_{s\gamma}T+(\lambda_0^{d+1})''\partial_s T\partial_\gamma T)\int_o^\infty d\tau'\int_{\tau'}^\infty d\tau\int d\xi
\Big\langle \bar w^{\beta k}(0,0)({\mathcal C}^{\gamma s})(\xi,\tau ),
$$
which becomes 
for $\beta=k$:
$$\qquad \qquad \sum_{\ga=1}^d\partial_k[((\lambda_0^{d+1})'\partial^2_{\gamma\gamma}T+(\lambda_0^{d+1})''\partial_\gamma T\partial_\gamma T)g_1],
$$
with
$$
 g_1:={1\over d}\sum_{\ga=1}^d\int_0^\infty d\tau'\int_{\tau'}^\infty d\tau\int d\xi
\Big\langle \bar w^{\beta k}(0,0)({\mathcal C}^{\gamma \gamma})(\xi,\tau)
$$

$$for \quad \beta\ne k\qquad \qquad \partial_k[((\lambda_0^{d+1})'\partial^2_{\beta k}T+(\lambda_0^{d+1})''\partial_k T\partial_\beta T)g_2],
$$
with
$$
g_2:=\int_0^\infty  d\tau'\int_{\tau'}^\infty d\tau\int d\xi
\Big\langle \bar w^{\beta k}(0,0)({\mathcal C}^{\beta\beta})(\xi,\tau ).
$$
and for the term ${\mathcal N}$: 
 $\e^{-d}\e\int dx  {\mathcal CN}^{\ s\gamma}(x)
\partial_s g_\ga( x) 
$ in the equation
for $\beta\ne k$ is zero and for $\beta=k$:
$$\qquad \qquad \sum_{\ga=1}^d\partial_k[((\lambda_0^{d+1})'\partial^2_{\gamma\gamma}T+(\lambda_0^{d+1})''\partial_\gamma T\partial_\gamma T)f_1],
$$
with
$$
 f_1:={1\over d}\sum_{\ga=1}^d\int_0^\infty d\tau'\int_{\tau'}^\infty d\tau\int d\xi
\Big\langle \bar w^{\beta k}(0,0)({\mathcal N}^{\gamma \gamma})(\xi,\tau)
$$

\vskip.1cm
\noindent ${\bf   Important \ remark}$

{\it This long calculation proves also that indeed $C$ is of order $1$ and hence the assumption on the definition of $R_1^s$ is correct. }
\vskip.1cm

The transport coefficients in \eqref{tau} in the stress thermal tensor $\tau^{(2)}$ have two different contributions. We can write $K_i=Y_i+X_i, i=1,2$. We have already given the expression of $Y_i$ and now we collect all the terms to give the expression of $X_i$.

$$X_1=\Big[[(\lambda_0^0)'(\lambda_0^{d+1})'] 4a_2(tz)+2b_2(\lambda_0^0)'(\lambda_0^{0})'+2c_2(\lambda_0^{d+1})'(\lambda_0^{d+1})'+(\lambda_0^{d+1})''[ h_2+2d_2+g_2]\Big ]\quad \beta\ne k,$$

$$
X_2=2(\lambda_0^{d+1})' d_2+2(\lambda_0^0)'T(z)h_2+(\lambda_0^{d+1})'g_2, \quad \beta\ne k.
$$
Moreover, there are also new contributions to $\omega_i$.  We write $\omega_i=\bar\omega_i+{\mathcal \phi}_i, i=1,2$. We get
$${\mathcal \phi}_1=[(\lambda_0^0)''h_1+(\lambda_0^{d+1})''d_1+a_1(\lambda_0^{d+1})'(\lambda_0^{0})'+b_1 (\lambda_0^0)'(\lambda_0^{0})'+c_1(\lambda_0^{d+1})'(\lambda_0^{d+1})'+(\lambda_0^{d+1})''(g_1+f_1),$$
$$
{\mathcal \phi}_2=(\lambda_0^0)'h_1+(\lambda_0^{d+1})'d_1+(\lambda_0^{d+1})'(g_1+f_1).
$$

\subsection{ Compatibility conditions}

We need to check that in the definitions of $R_i$ we can apply ${\mathcal L}^{-1}$ on the l.h.s.

$\bullet$ $R_1^a:\quad $ The simplest case is the definition of $R_1^a$ in \eqref{(2.31.2)}. We need to know that ${\mathcal L}^*G_0$ has zero projection on the null space. This is equivalent to prove 
$$\int dx z^\mu(x){\mathcal L}^*G_0=0,$$
since $Z^\mu=\int dx z^\mu(x)$ are the total mass, the total momentum and the total energy and they are conserved by the dynamics, namely ${\mathcal L}Z^\mu=0$. Hence for any observable $\phi$ $\int dx z^\mu(x){\mathcal L}^*\phi=0$. We have
$$\int dx z^\mu(x){\mathcal L}^*G_0=\int dx {\mathcal L} z^\mu(x)G_0=\sum_{\ga=1}^d{\partial_\gamma}\Big\langle \e^d\sum_i  \delta(x-x_i) w^{\mu  \gamma} _i]\Big\rangle_{G_0}.$$
For $\mu=0,d+1$, $\int dx z^\mu(x){\mathcal L}^*G_0$ is zero since $\Big\langle v_i\Big\rangle_{G_0}=0$. For  ${\mu=1\cdots d}$
$$\int dx z^\mu(x){\mathcal L}^*G_0=\sum_{\ga=1}^d{\partial_\gamma}\Big\langle\e^d\sum_i  w^{\mu  \gamma} _i\Big\rangle_{G_0}=\sum_{\ga=1}^d\delta_{\mu\ga}{\partial_\ga}\Big\langle \e^d\sum_i  w^{\mu  \mu} _i]\Big\rangle_{G_0}={\partial_\mu}P=0
$$
since $P$ is constant.
\vskip.1cm 

$\bullet$ $R_2:\quad $ $R_2$ is defined by \eqref{(2.33)}. It is not difficult to see that the condition in this case amount to say that $\rho$ and $e$ are solutions of the continuity equation and the energy equation. We have the condition to be verified for each $\mu$
$$0=\int dx z^\mu(x) [{\mathcal L}^*G_0g_1 +\partial_t G_0]=\sum_{k=1}^d\int dx {u^k\over T}\Big\langle{\mathcal L}z^\mu(x) z^k\Big\rangle_{G_0}+\int dx \partial_t\Big\langle z^\mu\Big\rangle_{G_0}.$$
This is naturally true for $\mu=1\cdots d$. Then,

$\mu=0:\quad $
\begin{equation}
\int dx \sum_{k=1}^d\partial_\mu[{u^k\over T}\Big\langle z^\mu z^k\Big\rangle_{G_0}]+\int dx \partial_t\Big\langle z^\mu\Big\rangle_{G_0}=\int dx [\partial_\mu(\rho u^\mu)+\partial_t \rho]=0.
\end{equation}
For $\mu=d+1$ we can add in the condition ${\mathcal L}^*G_0R_1$ since we have already proven that it is orthogonal to the invariants.

$\mu=d+1:\qquad \qquad  \qquad $
$$
\int dx z^{d+1}(x) [{\mathcal L}^*G_0g_1 +\partial_t G_0+{\mathcal L}^*G_0R_1]=
$$
$$
\int dx \sum_{k=1}^d\partial_\nu[{u^k\over T}\Big\langle w^{d+1\nu} z^k\Big\rangle_{G_0}]+\int dx \partial_t\Big\langle z^{d+1}\Big\rangle_{G_0}+\int dx \Big\langle w^{d+1\nu}G_0R_1\Big\rangle_{G_0}
$$
$$=\int dx [\partial_\mu((\rho e+P)u^\mu)+\partial_t (\rho e)-\nabla(k \nabla T )]=0.$$
\vskip.1cm 
$\bullet$ $R_1^s:\quad $ The most difficult case is   the definition of $R_1^s$ in \eqref{(2.31.1)}.
We need to show that ${\mathcal L}^*{\mathcal L}^*G_0$ has zero projection on the null.
We can write
$${\mathcal L}^*{\mathcal L}^*G_0={\mathcal L}^*[G_0\sum_{i=1}^N \sum_{\ga=1}^d\sum_{\mu=0,4}{\partial_\gamma \lambda_0 ^\mu} (x_i) w^{\mu  \gamma} _i]=\e^{-d}{\mathcal L}^*[\sum_{\mu=0,4}\sum_{\ga=1}^d\int dy G_0{\partial_\gamma \lambda_0 ^\mu } (y) w^{\mu  \gamma} (y)]\Big\rangle.$$
The condition to be satisfied is
$$\e^{-d}\int dx \Big\langle z^\a(x){\mathcal L}^*[\sum_{\mu=0,4}\sum_{\ga=1}^d\int dy G_0{\partial_\gamma \lambda_0 ^\mu} (y) w^{\mu  \gamma} (y)]\Big\rangle=0,\quad \ for \ \alpha=0,\cdots, d+1.$$
The lhs is
$$
\e^{-d}\int dx \Big\langle {\mathcal L} z^\a(x)[\sum_{\mu=0,4}\sum_{\ga=1}^d\int dy G_0{\partial_\gamma \lambda_0 ^\mu 
} (y) w^{\mu  \gamma} (y)]\Big\rangle
$$
$$
=\e^{-d}\int dx \Big\langle \nabla\cdot w^\a(x)[\sum_{\mu=0,4}\sum_{\ga=1}^d\int dy G_0{\partial_\gamma \lambda_0 ^\mu 
}  (y) w^{\mu  \gamma} (y)]\Big\rangle
$$
$$
=\e^{-d}\int dx \sum_{\nu,\ga=1}^d\sum_{\mu=0,4}\partial_\nu\int dy {\partial_\gamma \lambda_0 ^\mu 
}  (y)\Big\langle w^{\a\nu}(x) w^{\mu  \gamma} (\xi)\Big\rangle_{G_0}
$$
$$
=\sum_{\nu,\ga=1}^d\sum_{\mu=0,4}\partial_\nu\int dy {\partial_\gamma \lambda_0 ^\mu 
}  (y)\int d\xi\Big\langle w^{\a\nu}(0) w^{\mu  \gamma} (\xi)\Big\rangle_{G_0}.
$$
We know that \cite{S}
$$
\int d\xi\Big\langle w^{\a\nu}(0) w^{\mu  \gamma} (\xi)\Big\rangle_{G_0}=0 \quad  if \ \alpha=1\cdots d,\quad \mu=0,d+1;\quad \int d\xi\Big\langle w^{0\nu}(0) w^{0  \gamma} (\xi)\Big\rangle_{G_0}=\rho  T\delta_{\nu\gamma};
$$
$$ 
\int d\xi\Big\langle w^{d+1\nu}(0) w^{d+1  \gamma} (\xi)\Big\rangle_{G_0}={T(\rho e+P)^2\over \rho}\delta_{\nu\gamma};\quad \int d\xi\Big\langle w^{0\nu}(0) w^{d+1  \gamma} (\xi)\Big\rangle_{G_0}=T(\rho e+P)\delta_{\nu\gamma},
$$
so that we have two conditions:
$$\int dy[ {\partial_\gamma \lambda_0 ^0 
} (y)\rho T(y)+{\partial_\ga \lambda_0 ^{d+1}}T(\rho e+P)]=0,
$$
$$\int dy[ {\partial_\gamma \lambda_0 ^0} (y)T(\rho e+P)(y)+{\partial_\gamma \lambda_0 ^{d+1}
}{T(e+P)^2\over\rho}]=0.$$
Now we prove that 
$${\partial_\gamma \lambda_0 ^0 }(y)\rho(y)+{\partial_\gamma \lambda_0 ^{d+1}}(\rho e+P)=0,$$
so that both conditions are satisfied.
$$\rho\partial_\gamma \log z-P\partial_\gamma\beta=$$
$$=\rho{\delta(\log z)\over \delta P}\partial_\gamma P+\rho{\delta(\log z)\over \delta \beta}\partial_\gamma \beta+\log z{\delta \rho\over \delta P}\partial_\gamma P-(\rho e+P)\partial_\gamma\beta.$$
Since $P$ is constant we are left with
$$
[\rho{\delta(\log z)\over \delta \beta}-(\rho e+P)]=0
$$ 
by thermodynamic relations.

\vskip.5cm

\subsection {Galileian invariance}

 We write \eqref{(GI)}+second term in \eqref{(3.18.1)} as

\begin{align}
\begin{split}
&\sum_{\mu=1}^d\sum_{\ga=1}^d \int dy\Big [\la_1^\mu(y) {\partial_\gamma \lambda_0 ^0
} (y)\int d\xi \Big\langle z(\xi) ^\mu w^{0  \gamma} (\xi){\mathcal{L} } ^{-1} \bar w^{ \be k}(0)
\Big\rangle_{G_0}\\
+&\sum_{\mu=1}^d\sum_{\ga=1}^d\int dy\Big [\la_1^\mu(y) {\partial_\gamma \lambda_0 ^{d+1}
} (y)\int d\xi \Big\langle z(\xi) ^\mu w^{d+1 \gamma} (\xi){\mathcal{L} } ^{-1} \bar w^{ \be k}(0)
\Big\rangle_{G_0}
\\
&-\sum_{\mu=1}^d\sum_{\ell=1}^du ^\mu {1\over T^2}\int dy{\partial_l T } (y)]\int d\xi 
\Big\langle \bar w^{\mu l}(\xi) {\mathcal{L} } ^{-1} \bar w^{ \be k}(0)
(0)\Big\rangle_{G_0} \Big ]=0\label{new}
\end{split}
\end{align}
and we want to prove that \eqref{new} is zero.

We start from the identity, for any $\ga,\be,k=1,\cdots,d$,
$${\mathcal D}:= \int d\xi\Big\langle  w^{0  \gamma} (\xi){\mathcal{L} } ^{-1} \bar w^{ \be k}(0)
\Big\rangle_{G_0}=0$$
and define ${\mathcal D}_s$ and ${\mathcal D}^1_s$ below the same quantity where all the velocity $v$ are changed to $v-s u$. We have also
$${d\over ds}{\mathcal D}_s|{_{s=0}}=0.$$
We compute the derivative
$${d\over ds}{\mathcal D}_s=\int d\xi\Big\langle {G_0} {d\over ds}(w^{0  \gamma}) (\xi){\mathcal{L} } ^{-1} \bar w^{ \be k}(0)
\Big\rangle+\int d\xi\Big\langle {d\over ds}{G_0} (w^{0  \gamma}) (\xi){\mathcal{L} } ^{-1} \bar w^{ \be k}(0)
\Big\rangle
$$
$$+\int d\xi\Big\langle  w^{0  \gamma} (\xi){d\over ds}[{\mathcal{L} } ^{-1}] \bar w^{ \be k}(0)+\int d\xi\Big\langle  w^{0  \gamma} (\xi){\mathcal{L} } ^{-1} [{d\over ds}\bar w^{ \be k}(0)].
$$
We use the identity
$${d\over ds}[{\mathcal{L} } ^{-1}]=-{\mathcal{L} } ^{-1}{d\over ds}[{\mathcal{L}}] {\mathcal{L} } ^{-1}.
$$
It is not difficult to see that the only surviving term evaluated in $s=0$ is
$$\int d\xi\Big\langle {d\over ds}{G_0} (w^{0  \gamma}) (\xi){\mathcal{L} } ^{-1} \bar w^{ \be k}(0)
\Big\rangle=-\sum_{\mu=1}^d{u^\mu\over T}\int d\xi\Big\langle {G_0} (z^\mu w^{0  \gamma}) (\xi){\mathcal{L} } ^{-1} \bar w^{ \be k}(0)
\Big\rangle$$
and this implies that the first term in \eqref{new} is  zero.

Now we start from the identity for any $\ga,\be,k=1,\cdots,d$
$${\mathcal D}^1:= \int d\xi\Big\langle  w^{d+1  \gamma} (\xi){\mathcal{L} } ^{-1} \bar w^{ \be k}(0)
\Big\rangle_{G_0}=0$$
and
$${d\over ds}{\mathcal D^1}_s|{_{s=0}}=0.$$
We have
\begin{align}
\begin{split}{d\over ds}{\mathcal D^1}_s|{_{s=0}}&=-\sum_{\mu=1}^du^\mu\int d\xi\Big\langle {G_0} (z^\mu w^{d+1  \gamma}) (\xi){\mathcal{L} } ^{-1} \bar w^{ \be k}(0)
\Big\rangle+\Big\langle {G_0} {d\over ds}(w^{d+1  \gamma}) (\xi){\mathcal{L} } ^{-1} \bar w^{ \be k}(0)
\Big\rangle\\
&=\sum_{\mu=1}^d\Big[-{u^\mu\over T}\int d\xi\Big\langle {G_0} (z^\mu w^{d+1  \gamma}) (\xi){\mathcal{L} } ^{-1} \bar w^{ \be k}(0)
\Big\rangle-u^\mu\Big\langle {G_0} (w^{\mu  \gamma}) (\xi){\mathcal{L} } ^{-1} \bar w^{ \be k}(0)
\Big\rangle\Big]=0.
\label{zero}
\end{split}
\end{align}
The sum of the last two terms in \eqref{new} is
\begin{align}
&\sum_{\mu=1}^d \Big [-\int dy{u^\mu\over T}(y) {1\over T^2}
{\partial_\gamma T}  (y)\int d\xi \Big\langle z(\xi) ^\mu w^{d+1 \gamma} (\xi){\mathcal{L} } ^{-1} \bar w^{ \be k}(0)
\Big\rangle_{G_0}
\\
&-u ^\mu {1\over T^2}\int dy{\partial_l T } (y)]\int d\xi 
\Big\langle \bar w^{\mu l}(\xi) {\mathcal{L} } ^{-1} \bar w^{ \be k}(0)
(0)\Big\rangle_{G_0} \Big ]
\end{align}
and is equal to zero by using the identity \eqref{zero}.

\end{document}